\documentclass[aps,prd,floats,floatfix,superscriptaddress,nofootinbib,longbibliography]{revtex4-1}
\pdfoutput=1
\usepackage{graphicx}
\usepackage{amssymb}
\usepackage{hyperref}

\usepackage{amsmath}
\usepackage{xcolor}

\def\figureautorefname~#1\null{Fig.\,#1\null}
\def\tableautorefname~#1\null{Tab.\,#1\null}

\def\equationautorefname~#1\null{Eq.\,(#1)\null}
\begin{document}

\title{Updated constraints on the Georgi-Machacek model from LHC Run 2}

\author{Ameen Ismail}
\email{ai279@cornell.edu}
\affiliation{Ottawa-Carleton Institute for Physics, Carleton University, 1125 Colonel By Drive, Ottawa, Ontario K1S 5B6, Canada}
\affiliation{Department of Physics, LEPP, Cornell University, Ithaca, New York 14853, USA}

\author{Heather E.~Logan}
\email{logan@physics.carleton.ca} 
\affiliation{Ottawa-Carleton Institute for Physics, Carleton University, 1125 Colonel By Drive, Ottawa, Ontario K1S 5B6, Canada}

\author{Yongcheng Wu}
\email{ycwu@physics.carleton.ca}
\affiliation{Ottawa-Carleton Institute for Physics, Carleton University, 1125 Colonel By Drive, Ottawa, Ontario K1S 5B6, Canada}

\date{March 3, 2020}                                 

\begin{abstract}
We study the experimental constraints on the Georgi-Machacek model from direct searches for the new Higgs bosons and measurements of the properties of the discovered 125~GeV Higgs boson.  We apply these by interfacing the public codes HiggsBounds~5.2.0 and HiggsSignals~2.2.1, which implement a large collection of direct-search constraints on extra Higgs bosons and measurements of the properties of the 125~GeV state to the model calculator GMCALC.  We also implement constraints involving searches for doubly-charged Higgs bosons and Drell-Yan production of a neutral Higgs with decays to photon pairs directly into GMCALC; these channels are not included in HiggsBounds but provide important constraints on the model, especially at low mass.  We identify new constraints on the model from $H_3^0 \to Z h$ and $H \to hh$ and point out that these channels remain promising.  We also determine the allowed range of couplings of the SM-like Higgs boson after all experimental constraints are applied and show that the direct searches for additional Higgs bosons are generally more constraining than the measurements of the 125~GeV Higgs boson couplings.  The updated GMCALC code will be released as version 1.5.0.
\end{abstract}

\maketitle

\section{Introduction}

The primary objective of the CERN Large Hadron Collider (LHC) is, arguably, the elucidation of the mechanism of electroweak symmetry breaking.  This entails a comprehensive study of the properties of the 125~GeV Higgs boson discovered in 2012~\cite{Aad:2012tfa,Chatrchyan:2012xdj}, as well as searches for additional Higgs bosons that can appear in extensions of the Standard Model (SM) Higgs sector.  As these measurements become more constraining and the number of such searches grows, it becomes increasingly important to apply them in a systematic way to models with extended Higgs sectors in order to maintain a realistic understanding of the remaining possibilities for new Higgs physics.

This task is made manageable by the public implementations of Higgs search and measurement results in packages such as HiggsBounds~\cite{Bechtle:2013wla}, HiggsSignals~\cite{Bechtle:2013xfa}, and HEPfit~\cite{deBlas:2019okz}.  It is made even easier when these packages are called directly from model calculators, as HiggsBounds/HiggsSignals is from the two-Higgs-doublet model (2HDM) calculator 2HDMC~\cite{Eriksson:2009ws}, or when extended Higgs models are implemented into global fitting packages, as the 2HDM and Georgi-Machacek (GM) models are in HEPfit~\cite{Cacchio:2016qyh,Chowdhury:2017aav,Chiang:2018cgb}.

The GM model~\cite{Georgi:1985nv,Chanowitz:1985ug} is a prototype for extensions of the SM Higgs sector by scalars in triplet or larger isospin representations, implemented in a way that preserves the value of the electroweak $\rho$ parameter at tree level.  Its interesting features include doubly- and singly-charged scalars that couple at tree level to vector boson pairs, as well as the possibility that the SM-like Higgs boson's couplings to $WW$ and $ZZ$ could be larger than that in the SM.  The most important direct-search constraints on the GM model considered up to now are from vector boson fusion (VBF) production of the doubly-charged Higgs with decays to like-sign $W$ boson pairs~\cite{Chiang:2014bia,Khachatryan:2014sta,Logan:2017jpr} and vector boson fusion production of the singly-charged Higgs that decays to $WZ$~\cite{Aad:2015nfa,Sirunyan:2017sbn,Aaboud:2018ohp}.  Drell-Yan production of the doubly-charged Higgs with decays to like-sign dimuons also constrains masses below about 76~GeV~\cite{Kanemura:2014ipa,Logan:2015xpa}, and the fermiophobic neutral scalar can be constrained at low mass by its diphoton decays~\cite{Delgado:2016arn,Degrande:2017naf}.  The fermiophilic charged Higgs boson $H_3^+$ can also be searched for through its decays to $t\bar b$~\cite{Ghosh:2019qie,Aaboud:2018cwk,CMS:2019yat}.

In this paper we report on our implementation of an interface between the GM model calculator GMCALC~\cite{Hartling:2014xma} and HiggsBounds/HiggsSignals, through which we evalute experimental constraints on the model. We also implement constraints directly in GMCALC for some search channels that are not included in HiggsBounds, including the aforementioned VBF production of the doubly-charged Higgs boson decaying to $W$ boson pairs.  We show that the most stringent constraints on the model parameter space come from these directly-implemented constraints, in particular VBF production of the doubly-charged Higgs decaying to like-sign $W$ bosons and Drell-Yan production of the neutral Higgs boson $H_5^0$ of the custodial fiveplet decaying to diphotons.  However, our implementation of HiggsBounds shows that decays of the CP-odd neutral Higgs $H_3^0$ to $Zh$ and of the CP-even custodial-singlet $H$ to $hh$, where $h$ is the 125~GeV Higgs boson, exclude small regions of parameter space that are not otherwise constrained and are therefore promising channels for future searches.  Through our implementation of HiggsSignals we show that the current measurements of the couplings of the 125~GeV Higgs boson $h$ do not provide substantial additional constraints on the parameter space.  Indeed, the direct searches for additional Higgs bosons constrain the couplings of $h$ to vector bosons (fermions) to lie within $\pm 20\%$ ($30\%$) of their SM values.

This paper is organized as follows.  In the next section we describe the GM model and define the benchmark in which some of our scans will be performed.
In~\autoref{sec:codes} we describe the GM model calculator GMCALC and our implementation of the calls to HiggsBounds~5.2.0 and HiggsSignals~2.2.1.  This implementation will be released in GMCALC 1.5.0.  In~\autoref{sec:constraints} we describe how we implemented the constraints from VBF and Drell-Yan production of the doubly-charged Higgs, as well as Drell-Yan production of $H_5^0$ with decays to diphotons. In~\autoref{sec:HB5} we show the constraints on the GM model from HiggsBounds.  We show results in two specific benchmarks as well as in general parameter scans at high and low mass.  In~\autoref{sec:HS} we apply HiggsSignals to constrain the properties of the 125~GeV SM-like Higgs boson and compare the resulting constraints to those from direct searches.  In~\autoref{sec:conclusions} we conclude.

\section{The Georgi-Machacek model}
\label{sec:model}

The scalar sector of the GM model~\cite{Georgi:1985nv,Chanowitz:1985ug} consists of the usual complex doublet $(\phi^+,\phi^0)$ of the SM with hypercharge\footnote{We use $Q = T^3 + Y/2$.} $Y = 1$, a real triplet $(\xi^+,\xi^0,\xi^-)$ with $Y = 0$, and  a complex triplet $(\chi^{++},\chi^+,\chi^0)$ with $Y=2$.  The doublet is responsible for the fermion masses as in the SM.
In order to make the global SU(2)$_L \times$SU(2)$_R$ symmetry explicit, we write the doublet in the form of a bidoublet $\Phi$ and combine the triplets to form a bitriplet $X$:
\begin{equation}
	\Phi = \left( \begin{array}{cc}
	\phi^{0*} &\phi^+  \\
	-\phi^{+*} & \phi^0  \end{array} \right), \qquad
	X =
	\left(
	\begin{array}{ccc}
	\chi^{0*} & \xi^+ & \chi^{++} \\
	 -\chi^{+*} & \xi^{0} & \chi^+ \\
	 \chi^{++*} & -\xi^{+*} & \chi^0  
	\end{array}
	\right).
	\label{eq:PX}
\end{equation}
The vevs are defined by $\langle \Phi  \rangle = \frac{ v_{\phi}}{\sqrt{2}} I_{2\times2}$  and $\langle X \rangle = v_{\chi} I_{3 \times 3}$, where $I$ is the appropriate identity matrix and the $W$ and $Z$ boson masses constrain
\begin{equation}
	v_{\phi}^2 + 8 v_{\chi}^2 \equiv v^2 = \frac{1}{\sqrt{2} G_F} \approx (246~{\rm GeV})^2,
	\label{eq:vevrelation}
\end{equation} 
where $G_F$ is the Fermi constant.
Upon electroweak symmetry breaking, the global SU(2)$_L \times $SU(2)$_R$ symmetry breaks down to the diagonal subgroup, which is the custodial SU(2) symmetry.

The most general gauge-invariant scalar potential involving these fields that conserves custodial SU(2) is given, in the conventions of Ref.~\cite{Hartling:2014zca}, by\footnote{A translation table to other parameterizations in the literature has been given in the appendix of Ref.~\cite{Hartling:2014zca}.}
\begin{eqnarray}
	V(\Phi,X) &= & \frac{\mu_2^2}{2}  \text{Tr}(\Phi^\dagger \Phi) 
	+  \frac{\mu_3^2}{2}  \text{Tr}(X^\dagger X)  
	+ \lambda_1 [\text{Tr}(\Phi^\dagger \Phi)]^2  
	+ \lambda_2 \text{Tr}(\Phi^\dagger \Phi) \text{Tr}(X^\dagger X)   \nonumber \\
          & & + \lambda_3 \text{Tr}(X^\dagger X X^\dagger X)  
          + \lambda_4 [\text{Tr}(X^\dagger X)]^2 
           - \lambda_5 \text{Tr}( \Phi^\dagger \tau^a \Phi \tau^b) \text{Tr}( X^\dagger t^a X t^b) 
           \nonumber \\
           & & - M_1 \text{Tr}(\Phi^\dagger \tau^a \Phi \tau^b)(U X U^\dagger)_{ab}  
           -  M_2 \text{Tr}(X^\dagger t^a X t^b)(U X U^\dagger)_{ab}.
           \label{eq:potential}
\end{eqnarray} 
Here the SU(2) generators for the doublet representation are $\tau^a = \sigma^a/2$ with $\sigma^a$ being the Pauli matrices, the generators for the triplet representation are
\begin{equation}
	t^1= \frac{1}{\sqrt{2}} \left( \begin{array}{ccc}
	 0 & 1  & 0  \\
	  1 & 0  & 1  \\
	  0 & 1  & 0 \end{array} \right), \qquad  
	  t^2= \frac{1}{\sqrt{2}} \left( \begin{array}{ccc}
	 0 & -i  & 0  \\
	  i & 0  & -i  \\
	  0 & i  & 0 \end{array} \right), \qquad 
	t^3= \left( \begin{array}{ccc}
	 1 & 0  & 0  \\
	  0 & 0  & 0  \\
	  0 & 0 & -1 \end{array} \right),
\end{equation}
and the matrix $U$, which rotates $X$ into the Cartesian basis, is given by~\cite{Aoki:2007ah}
\begin{equation}
	 U = \left( \begin{array}{ccc}
	- \frac{1}{\sqrt{2}} & 0 &  \frac{1}{\sqrt{2}} \\
	 - \frac{i}{\sqrt{2}} & 0  &   - \frac{i}{\sqrt{2}} \\
	   0 & 1 & 0 \end{array} \right).
	 \label{eq:U}
\end{equation}

The minimization conditions for the scalar potential read
\begin{eqnarray}
	0 = \frac{\partial V}{\partial v_{\phi}} &=& 
	v_{\phi} \left[ \mu_2^2 + 4 \lambda_1 v_{\phi}^2 
	+ 3 \left( 2 \lambda_2 - \lambda_5 \right) v_{\chi}^2 - \frac{3}{2} M_1 v_{\chi} \right], 
		\nonumber \\
	0 = \frac{\partial V}{\partial v_{\chi}} &=& 
	3 \mu_3^2 v_{\chi} + 3 \left( 2 \lambda_2 - \lambda_5 \right) v_{\phi}^2 v_{\chi}
	+ 12 \left( \lambda_3 + 3 \lambda_4 \right) v_{\chi}^3
	- \frac{3}{4} M_1 v_{\phi}^2 - 18 M_2 v_{\chi}^2.
	\label{eq:mincond}
\end{eqnarray}

The physical fields can be organized by their transformation properties under the custodial SU(2) symmetry into a fiveplet, a triplet, and two singlets.  The fiveplet and triplet states are given by
\begin{eqnarray}
	&&H_5^{++} = \chi^{++}, \qquad
	H_5^+ = \frac{\left(\chi^+ - \xi^+\right)}{\sqrt{2}}, \qquad
	H_5^0 = \sqrt{\frac{2}{3}} \xi^{0,r} - \sqrt{\frac{1}{3}} \chi^{0,r}, \nonumber \\
	&&H_3^+ = - s_H \phi^+ + c_H \frac{\left(\chi^++\xi^+\right)}{\sqrt{2}}, \qquad
	H_3^0 = - s_H \phi^{0,i} + c_H \chi^{0,i},
\end{eqnarray}
where we have decomposed the neutral fields into real and imaginary parts according to
\begin{equation}
	\phi^0 \to \frac{v_{\phi}}{\sqrt{2}} + \frac{\phi^{0,r} + i \phi^{0,i}}{\sqrt{2}},
	\qquad
	\chi^0 \to v_{\chi} + \frac{\chi^{0,r} + i \chi^{0,i}}{\sqrt{2}}, 
	\qquad
	\xi^0 \to v_{\chi} + \xi^{0,r}.
	\label{eq:decomposition}
\end{equation}
and the vevs are parameterized by
\begin{equation}
	c_H \equiv \cos\theta_H = \frac{v_{\phi}}{v}, \qquad
	s_H \equiv \sin\theta_H = \frac{2\sqrt{2}\,v_\chi}{v}.
\end{equation}
The parameter $s_H^2$ has an interesting physical meaning, being the fraction of $M_W^2$ and $M_Z^2$ that is generated by the triplet vevs at tree level.

The masses within each custodial multiplet are degenerate at tree level and can be written (after eliminating $\mu_2^2$ and $\mu_3^2$ in favor of the vevs) as\footnote{Note that the ratio $M_1/v_{\chi}$ is finite in the limit $v_{\chi} \to 0$, 
\begin{equation}
	\frac{M_1}{v_{\chi}} = \frac{4}{v_{\phi}^2} 
	\left[ \mu_3^2 + (2 \lambda_2 - \lambda_5) v_{\phi}^2 
	+ 4(\lambda_3 + 3 \lambda_4) v_{\chi}^2 - 6 M_2 v_{\chi} \right],
\end{equation}
which follows from the minimization condition $\partial V/\partial v_{\chi} = 0$.}
\begin{eqnarray}
	m_5^2 &=& \frac{M_1}{4 v_{\chi}} v_\phi^2 + 12 M_2 v_{\chi} 
	+ \frac{3}{2} \lambda_5 v_{\phi}^2 + 8 \lambda_3 v_{\chi}^2, \nonumber \\
	m_3^2 &=&  \frac{M_1}{4 v_{\chi}} (v_\phi^2 + 8 v_{\chi}^2) 
	+ \frac{\lambda_5}{2} (v_{\phi}^2 + 8 v_{\chi}^2) 
	= \left(  \frac{M_1}{4 v_{\chi}} + \frac{\lambda_5}{2} \right) v^2.
\end{eqnarray}

The two custodial SU(2)--singlet mass eigenstates are given by
\begin{equation}
	h = \cos \alpha \, \phi^{0,r} - \sin \alpha \, H_1^{0\prime},  \qquad
	H = \sin \alpha \, \phi^{0,r} + \cos \alpha \, H_1^{0\prime},
	\label{mh-mH}
\end{equation}
where 
\begin{equation}
	H_1^{0 \prime} = \sqrt{\frac{1}{3}} \xi^{0,r} + \sqrt{\frac{2}{3}} \chi^{0,r}.
\end{equation}
The elements of their mass matrix in the basis $(\phi^{0,r}, H_1^{0\prime})$ are given by
\begin{eqnarray}
	\mathcal{M}_{11}^2 &=& 8 \lambda_1 v_{\phi}^2, \nonumber \\
	\mathcal{M}_{12}^2 &=& \frac{\sqrt{3}}{2} v_{\phi} 
	\left[ - M_1 + 4 \left(2 \lambda_2 - \lambda_5 \right) v_{\chi} \right], \nonumber \\
	\mathcal{M}_{22}^2 &=& \frac{M_1 v_{\phi}^2}{4 v_{\chi}} - 6 M_2 v_{\chi} 
	+ 8 \left( \lambda_3 + 3 \lambda_4 \right) v_{\chi}^2.
\end{eqnarray}
In this paper we break with tradition and define $h$ to be the 125~GeV Higgs, and $H$ to be the other custodial-singlet state, which can be heavier or lighter than $h$.

The custodial-fiveplet states $H_5^{0,\pm,\pm\pm}$ are composed entirely of isospin triplet scalars, and hence are fermiophobic.\footnote{We neglect the possible lepton-number-violating coupling of the isospin triplet $\chi$ to two lepton doublets, the strength of which is of order $m_{\nu}/v_{\chi} \sim 10^{-12}/s_H$.}  The fiveplet states do, however, couple at tree level to vector boson pairs with a coupling proportional to $s_H$.  The custodial-triplet states $H_3^{0,\pm}$, on the other hand, couple to fermions with strength proportional to $s_H$ and do not couple to vector boson pairs at tree level -- their phenomenology is similar to that of the pseudoscalar and charged Higgs of the Type-I 2HDM.  The second custodial singlet $H$ couples to both vector boson pairs and fermion pairs at tree level.

The scalar potential of the GM model in~\autoref{eq:potential} contains 9 parameters, two of which can be fixed by the measured values of $G_F$ and $m_h$.  This leaves a 7-dimensional parameter space to be scanned over.  In addition to general scans over the full parameter space, it is also useful to consider strategically-chosen benchmark planes, which are two-dimensional slices through the parameter space.  Two benchmark planes have been proposed for the GM model: the so-called H5plane benchmark (\autoref{tab:H5plane}), which was introduced in Ref.~\cite{deFlorian:2016spz} and its phenomenology studied in some detail in Ref.~\cite{Logan:2017jpr}, and the low-$m_5$ benchmark (\autoref{tab:lowm5}), which was introduced in particular to cover the $m_5$ range below 200~GeV and is being studied in Ref.~\cite{Ismail:2020kqz}.

\begin{table}[!tbp]
    \begin{tabular}{lll}
    \hline \hline
    Fixed inputs & Variable parameters & Other parameters \\ \hline
    $G_F= 1.1663787 \times 10^{-5}~\rm{GeV}^{-2}$ & $m_5 \in [200,3000]~\rm{GeV} $ & $\lambda_2 =0.4 m_5 / (1000~{\rm GeV})$ \\
    $m_h = 125~{\rm GeV}$ & $s_H \in (0,1)$ & $M_1 = \sqrt{2} s_H(m^2_5 +v^2)/v$\\
    $\lambda_3 = -0.1$ &  & $M_2 = M_1/6$\\
    $\lambda_4 = 0.2$ &  & \\ \hline\hline
    \end{tabular}
\caption{Parameter definitions for the H5plane benchmark scenario~\cite{deFlorian:2016spz} in the GM model.}
\label{tab:H5plane}
\end{table}

\begin{table}[!tbp]
	\begin{tabular}{lll}
	\hline\hline
	Fixed inputs & Variable parameters & Other parameters \\ \hline
	$G_F = 1.1663787 \times 10^{-5}$~GeV$^{-2}$ & $m_5 \in (50, 550)$~GeV & $\lambda_2 = 0.08(m_5/100~{\rm GeV})$ \\
	$m_h = 125$~GeV & $s_H \in (0,1)$ & $\lambda_5 = -4 \lambda_2 = -0.32 (m_5/100~{\rm GeV})$ \\
	$\lambda_3 = -1.5$ & &  \\
	$\lambda_4 = -\lambda_3 = 1.5$& &  \\
	$M_2 = 10$~GeV & &  \\
	\hline\hline
	\end{tabular}
\caption{Parameter definitions for the low-$m_5$ benchmark scenario~\cite{Ismail:2020kqz} in the GM model.}
\label{tab:lowm5}
\end{table}

Both benchmarks take $m_5$ and $s_H$ as their two free parameters.  The rest of the parameters are chosen so that (i) the custodial-fiveplet states $H_5^{0, \pm, \pm\pm}$ are lighter than $H_3^{0,\pm}$ and $H$, thereby ensuring that $H_5^{\pm\pm}$ and $H_5^{\pm}$ decay exclusively into vector boson pairs, and (ii) the benchmark comes close to populating the full theoretically-allowed range of $m_5$ and $s_H$ accessible in a general scan.  These properties make these benchmarks suitable for interpreting LHC searches for $H_5^{\pm\pm}$ and $H_5^{\pm}$ produced in vector boson fusion (the cross section for which is proportional to $s_H^2$) and decaying into vector boson pairs, as well as for Drell-Yan production of $H_5^{++}H_5^{--}$ at lower masses.  We will evaluate constraints from HiggsBounds and HiggsSignals in these benchmarks as well as in general scans of the parameter space and compare their effectiveness to dedicated searches for doubly-charged Higgs bosons.

\section{The codes}
\label{sec:codes}

GMCALC~\cite{Hartling:2014xma} is a public Fortran code that, given a set of input parameters, calculates the particle spectrum, couplings, and decay widths of the scalars in the GM model.  It also implements checks of the theoretical constraints on the model parameters from perturbative unitarity of two-to-two scalar scattering amplitudes, boundedness-from-below of the scalar potential, and the absence of deeper minima following Ref.~\cite{Hartling:2014zca}.  The numerical calculations used in this analysis were based on GMCALC version 1.4.1.\footnote{GMCALC 1.4.1 includes the nontrivial one-loop calculations of $H_5^0 \to Z\gamma$ and $H_3^+, H_5^+ \to W^+ \gamma$~\cite{Degrande:2017naf}, for which the external LoopTools package~\cite{Hahn:1998yk} must be installed. In practice, only $H_5^0 \to Z \gamma$ matters for the results in this paper: it affects the branching ratio of $H_5^0 \to \gamma\gamma$ at masses below the $WW$ threshold, modifying it by up to 10\%~\cite{Degrande:2017naf}.  If precision at this level is not needed, the LoopTools calls can safely be omitted, in which case BR($H_5^0 \to Z\gamma$) and BR($H_3^+,H_5^+ \to W^+ \gamma$) are set to zero in the code.}

For this study, HiggsBounds was run in the effective coupling mode, which was most amenable to integration with GMCALC. HiggsSignals was run with the same inputs as HiggsBounds, employing a Gaussian probability distribution for the Higgs masses. HiggsSignals was used to constrain the 125 GeV Higgs, while HiggsBounds exclusions were applied to all other Higgses \footnote{It would not make sense to apply HiggsBounds to constrain the 125 GeV Higgs. Rather, we use HiggsSignals, to test the GM model predictions against measured signal strengths of the 125 GeV Higgs.}. The various coupling modification factors and branching ratios required as input were calculated from the output of GMCALC. All rate and mass uncertainties were set to zero. On rare occasion, the branching ratio for a loop-induced decay of some Higgs was nonzero in the GM model but zero for an SM Higgs of the same mass. In such cases the relevant coupling modification factor was simply set to zero. The total decay width of each neutral Higgs was recalculated using the coupling modification factors and SM Higgs branching ratios (determined using HiggsBounds internal functions), as opposed to using the value computed by GMCALC. This is required to ensure the total width is consistent with the sum of the partial widths.

Grid-like scans were performed over the H5plane benchmark, varying $m_5$ from $200$ GeV to $1050$ GeV in increments of $10$ GeV, and $s_H$ from 0 to 1 in increments of 0.01. The upper bound on $m_5$ is the maximum mass at which HiggsBounds can calculate SM Higgs branching ratios. A similar scan was performed over the low-$m_5$ benchmark, varying $m_5$ from $50$ GeV to $200$ GeV and $s_H$ from 0 to 1 (in the same increments). In the scans over the whole parameter space, points were sampled randomly, with bounds on parameters chosen in accordance with the maximum theoretically allowed ranges described in Ref.~\cite{Hartling:2014zca}. The $m_5$ ranges for these general scans were chosen to match those in the benchmarks.  For the H5plane and general scans, GMCALC was run with INPUTSET = 4, while in the low-$m_5$ benchmark scan we used INPUTSET = 6 (see \cite{Hartling:2014xma}). Both of these modes allow direct input of $m_5$ and $s_H$. The number of model parameters, which is required to compute $p$-values in HiggsSignals, was taken to be two in all scans. In the benchmark scans, there are in fact two free parameters, $m_5$ and $s_H$, and the same number was chosen for the general scans so that they could be compared to the benchmarks on an equal statistical footing.

\section{Constraints from doubly-charged Higgs bosons and Drell-Yan diphotons}
\label{sec:constraints}

HiggsBounds~5.2.0 does not include constraints from searches for doubly-charged Higgs bosons or from Drell-Yan production of a neutral Higgs boson decaying to $\gamma\gamma$.  These processes constitute some of the strongest direct constraints on the GM model.  We implement the following processes directly in GMCALC.  

\subsection{VBF $H_5^{\pm\pm} \to W^{\pm} W^{\pm} \to$ like-sign dileptons}

The current most sensitive search for vector boson fusion (VBF) production of $H_5^{\pm\pm}$ with decays to $W^{\pm}W^{\pm}$ for $m_5 \geq 200$~GeV is from a CMS analysis of 35.9~fb$^{-1}$ of LHC Run 2 (13~TeV) data~\cite{Sirunyan:2017ret}.  The upper bound on $s_H$ as a function of $m_5$ appears in the supplemental material of the published version and assumes BR($H_5^{++} \to W^+W^+) = 1$.  We take into account the possibility that BR($H_5^{++} \to W^+W^+) < 1$ when $m_3 < m_5$ by using the fact that the signal production cross section is proportional to $s_H^2$, so that
\begin{equation}
	(s_H^{\rm limit})^2 \times {\rm BR}(H_5^{++} \to W^+W^+) = (s_H^{\rm CMS})^2,
	\label{eq:BRlt1}
\end{equation}
where $s_H^{\rm CMS}$ is the limit from Ref.~\cite{Sirunyan:2017ret} for BR($H_5^{++} \to W^+W^+) = 1$.  

For $m_5 < 200$~GeV, VBF production of $H_5^{\pm\pm}$ with decays to $W^{\pm}W^{\pm}$ is constrained by an ATLAS measurement of the VBF like-sign $W$ boson production cross section using 20.3~fb$^{-1}$ of LHC Run 1 (8~TeV) data~\cite{Aad:2014zda}, which was recast in Ref.~\cite{Chiang:2014bia} to constrain $H_5^{\pm\pm}$ production in the GM model.\footnote{We thank Cheng-Wei Chiang for providing the numerical version of the exclusion contour of Ref.~\cite{Chiang:2014bia}.}  The recast puts an upper bound on $v_{\chi}$ (equivalently $s_H$) as a function of $m_5$ assuming BR($H_5^{++} \to W^+W^+) = 1$.  We account for the possibility that BR($H_5^{++} \to W^+W^+) < 1$ in the same way as~\autoref{eq:BRlt1}.  

\subsection{Drell-Yan $H_5^{\pm\pm}$ with $H_5^{\pm\pm} \to W^\pm W^\pm \to$ like-sign dileptons}

Drell-Yan production of $H_5^{++}H_5^{--}$ and $H_5^{\pm\pm} H_5^{\mp}$ with $H_5^{\pm\pm} \to W^{\pm} W^{\pm}$ is constrained by an ATLAS search for anomalous like-sign dimuon production using 20.3~fb$^{-1}$ of LHC Run 1 (8~TeV) data~\cite{ATLAS:2014kca}, which was recast in Ref.~\cite{Kanemura:2014ipa} to constrain the Higgs Triplet Model assuming degenerate $H^{++}$ and $H^+$.  The latter was reinterpreted in Ref.~\cite{Logan:2015xpa} in the GM model, assuming BR($H_5^{++} \to W^+W^+) = 1$; in this case, the measurement excludes $m_5$ values below about 76~GeV independent of $s_H$.  We take into account the possibility that BR($H_5^{++} \to W^+W^+) < 1$ by applying the upper limit on the fiducial cross section from Ref.~\cite{ATLAS:2014kca} to the quantity
\begin{eqnarray}
	\sigma_{\rm fiducial} &=& 0.95 \times \left[ \sigma_{H_5^{++}H_5^{--}} 
	\left( 2 \, {\rm BR}(H_5^{++} \to \mu^+ \mu^+) \epsilon_{H_5^{++}H_5^{--}} 
	- {\rm BR}(H_5^{++} \to \mu^+ \mu^+)^2 \epsilon_{H_5^{++}H_5^{--}}^2 \right) \right. \nonumber \\
	&& \left. + \sigma_{H_5^{++} H_5^-} {\rm BR}(H_5^{++} \to \mu^+ \mu^+) \epsilon_{H_5^{++}H_5^-}
	+ \sigma_{H_5^{--} H_5^+} {\rm BR}(H_5^{++} \to \mu^+ \mu^+) \epsilon_{H_5^{--}H_5^+}
	\right],
\end{eqnarray}
where ${\rm BR}(H_5^{++} \to \mu^+ \mu^+) = {\rm BR}(H_5^{++} \to W^+W^+) \times {\rm BR}(W^+W^+ \to \mu^+\mu^+ + {\rm MET})$.  We take the values of ${\rm BR}(W^+W^+ \to \mu^+\mu^+ + {\rm MET})$, the cross sections for the Higgs Triplet Model, and the efficiencies $\epsilon_{ij}$ from Ref.~\cite{Kanemura:2014ipa}.  The cross sections $\sigma_{ij}$ for the GM model are related to those in the Higgs Triplet Model (HTM) by~\cite{Logan:2015xpa}
\begin{equation}
	\sigma_{H_5^{++}H_5^{--}} = \sigma_{H^{++}H^{--}}^{\rm HTM}, \qquad
	\sigma_{H_5^{++}H_5^-} = \frac{1}{2} \sigma_{H^{++}H^-}^{\rm HTM}, \qquad
	\sigma_{H_5^{--}H_5^+} = \frac{1}{2} \sigma_{H^{--}H^+}^{\rm HTM}.
\end{equation} 
The values of ${\rm BR}(W^+W^+ \to \mu^+\mu^+ + {\rm MET})$ do not follow straightforwardly from the individual $W$ decay branching ratios because of quantum mechanical interference when $H_5^{++}$ is lighter than the $WW$ threshold.
The factor of 0.95 conservatively takes into account the $\pm 5\%$ theory uncertainty on the signal cross sections, computed at next-to-leading order in QCD.

\subsection{Drell-Yan $H_5^0 H_5^{\pm}$ with $H_5^0 \to \gamma\gamma$}

\begin{figure}[!btp]
  \centering
  \includegraphics[width=0.48\textwidth]{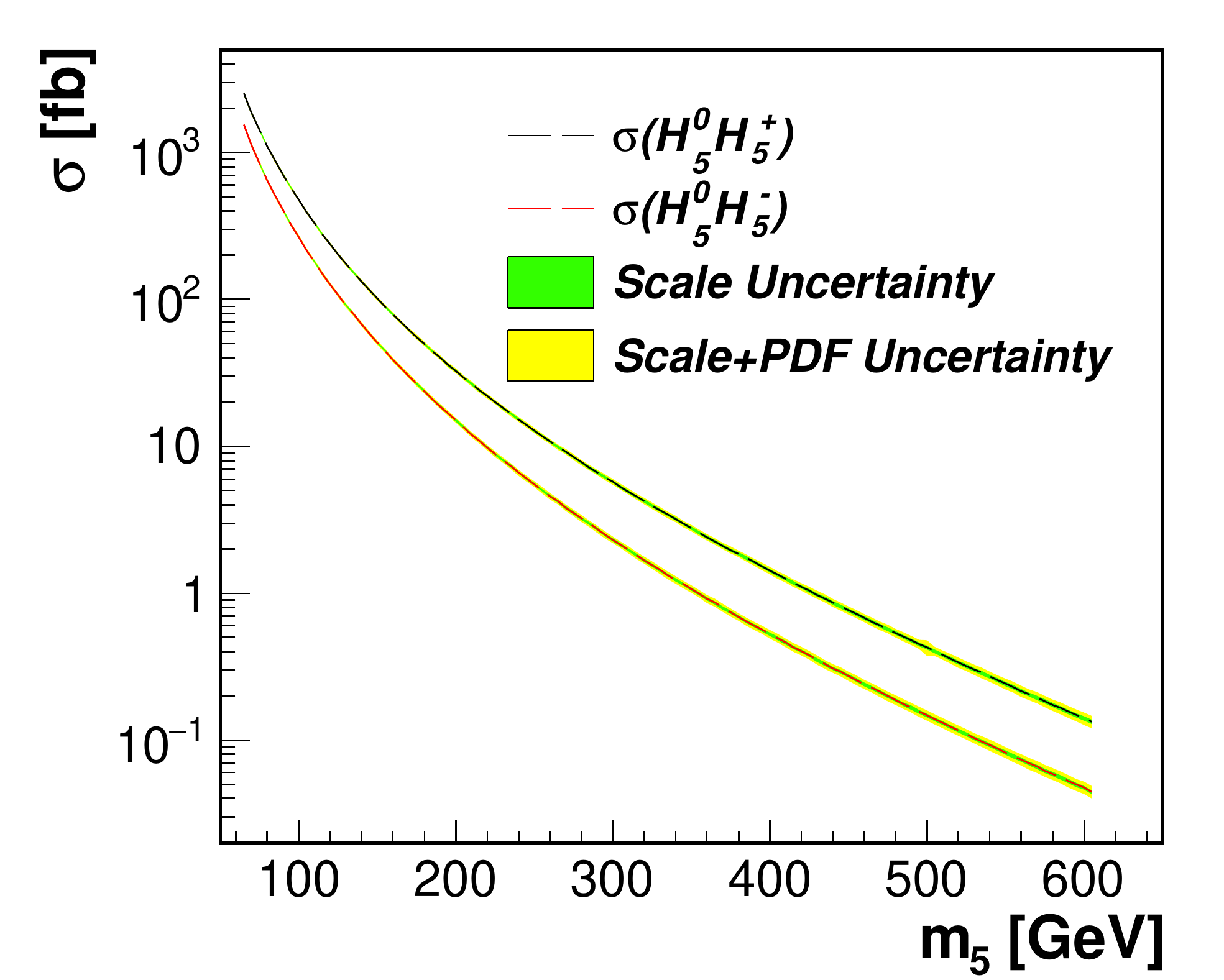}
  \put(-190,50){$\sqrt{s} = 8$ TeV}
  \includegraphics[width=0.48\textwidth]{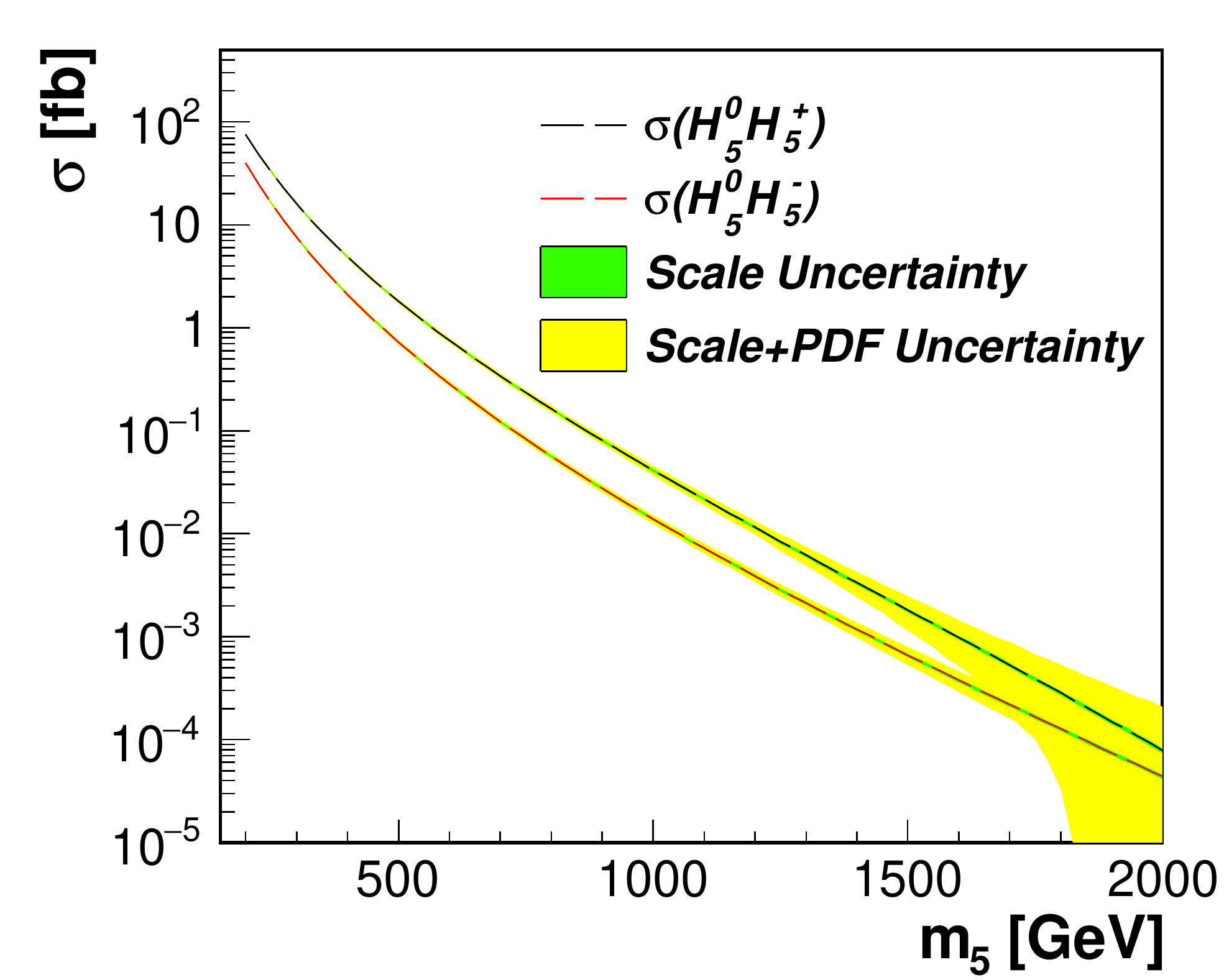}
  \put(-190,50){$\sqrt{s} = 13$ TeV}
  \caption{The cross sections of $H_5^0H_5^\pm$ at the LHC with $\sqrt{s}=8$ TeV (left panel) and $\sqrt{s}=13$ TeV (right panel), computed using MadGraph5 at NLO in QCD.}
  \label{fig:cs_h50h5pm_8_13_TeV}
\end{figure}

\begin{figure}[!btp]
  \centering
  \includegraphics[width=0.48\textwidth]{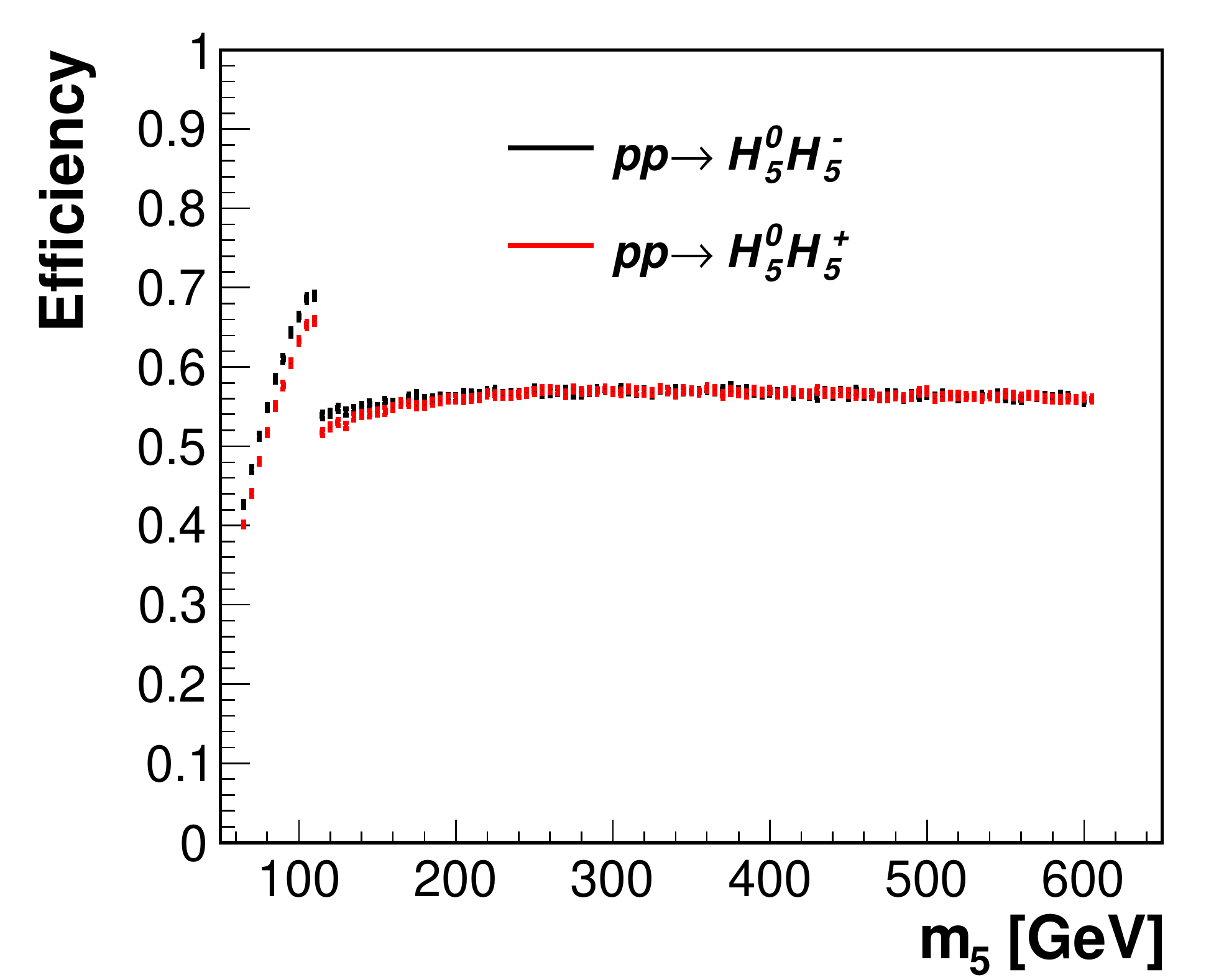}
  \put(-190,50){$\sqrt{s} = 8$ TeV}
  \includegraphics[width=0.48\textwidth]{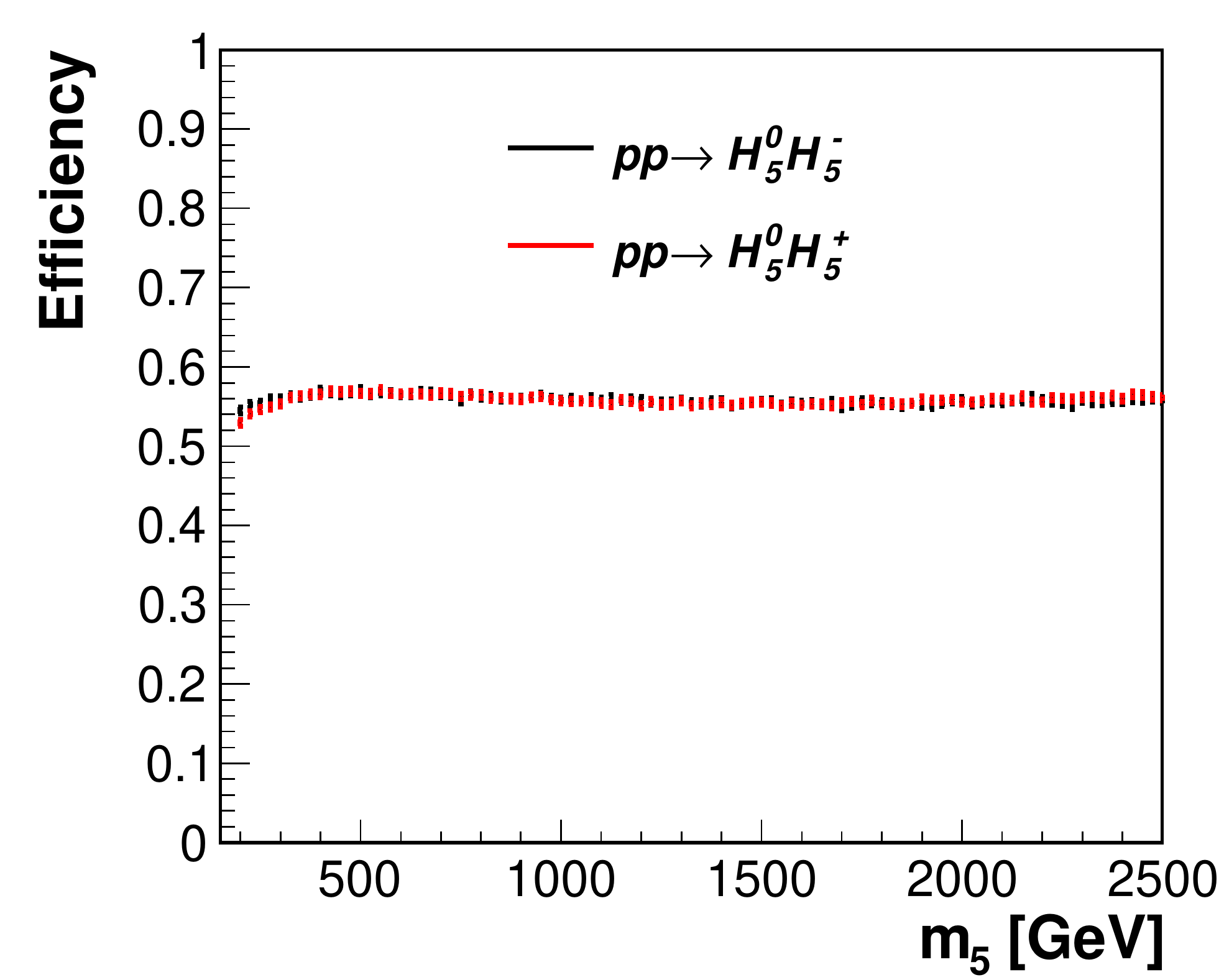}
  \put(-190,50){$\sqrt{s} = 13$ TeV}
  \caption{The efficiencies $\epsilon_{\pm}$ of the experimental selection cuts for $H_5^0(\to\gamma\gamma)H_5^\pm$ at the LHC with $\sqrt{s}=8$ TeV (left panel) and $\sqrt{s}=13$ TeV (right panel), computed using simulated events in MadGraph5 at NLO in QCD.}
  \label{fig:eff_h50h5pm_8_13_TeV}
\end{figure}

Drell-Yan production of $H_5^0 H_5^{\pm}$ with $H_5^0 \to \gamma\gamma$ is constrained by an ATLAS search for diphoton resonances in the mass range 65--600~GeV using 20.3~fb$^{-1}$ of LHC Run 1 (8~TeV) data~\cite{Aad:2014ioa} as well as in the mass range 200--2700~GeV using 36.7~fb$^{-1}$ of LHC Run 2 (13~TeV) data~\cite{Aaboud:2017yyg}. The constraints are placed on fiducial cross section times branching fraction. We first generate the cross sections and events at next-to-leading order (NLO) in QCD using MadGraph5~\cite{Alwall:2014hca} at 8~TeV and 13~TeV $pp$ centre-of-mass energies for $H_5^0H_5^+$ and $H_5^0H_5^-$ separately. The total cross sections for these two processes are shown in~\autoref{fig:cs_h50h5pm_8_13_TeV}. The fiducial cross section is obtained by applying the cuts used in Refs.~\cite{Aad:2014ioa} and~\cite{Aaboud:2017yyg} respectively:
\begin{description}
  \item[8~TeV]
  \ \  
  \begin{itemize}
    \item $|\eta_{\gamma}|<2.37$,
    \item For $m_{\gamma\gamma} > 110$ GeV, $p^T_{\gamma_1}>0.4 m_{\gamma\gamma}$ and $p^T_{\gamma_2}>0.3 m_{\gamma\gamma}$,
    \item For $m_{\gamma\gamma}<110$ GeV, $p^T_\gamma > 22$ GeV,
  \end{itemize} 
  \item[13~TeV]
  \ \  
  \begin{itemize}
    \item $|\eta_{\gamma}|<2.37$,
    \item $p^T_{\gamma_1}>0.4 m_{\gamma\gamma}$ and $p^T_{\gamma_2}>0.3 m_{\gamma\gamma}$,
  \end{itemize}
\end{description}
where $\eta_{\gamma}$ is the pseudorapidity of each of the two photons, $m_{\gamma\gamma}$ is the diphoton invariant mass, and $p^T_{\gamma_i}$ are the transverse momenta of each of the photons.
The corresponding efficiencies, $\epsilon_{\pm}$ for $H_5^0H_5^{\pm}$ respectively, in going from total cross section to fiducial cross section are shown in~\autoref{fig:eff_h50h5pm_8_13_TeV}.  The experimental upper limit on the fiducial cross section times branching ratio is then applied for each mass point $m_5$ to the quantity
\begin{align}
  \sigma_{\rm fiducial} = (\sigma_{H_5^0H_5^+}\times\epsilon_+ + \sigma_{H_5^0H_5^-}\times\epsilon_-)\times {\rm BR}(H_5^0\to\gamma\gamma).
\end{align}

\section{Constraints from HiggsBounds 5.2.0}
\label{sec:HB5}

\subsection{H5plane benchmark}
\begin{figure}[!tbp]
\resizebox{0.5\textwidth}{!}{\includegraphics{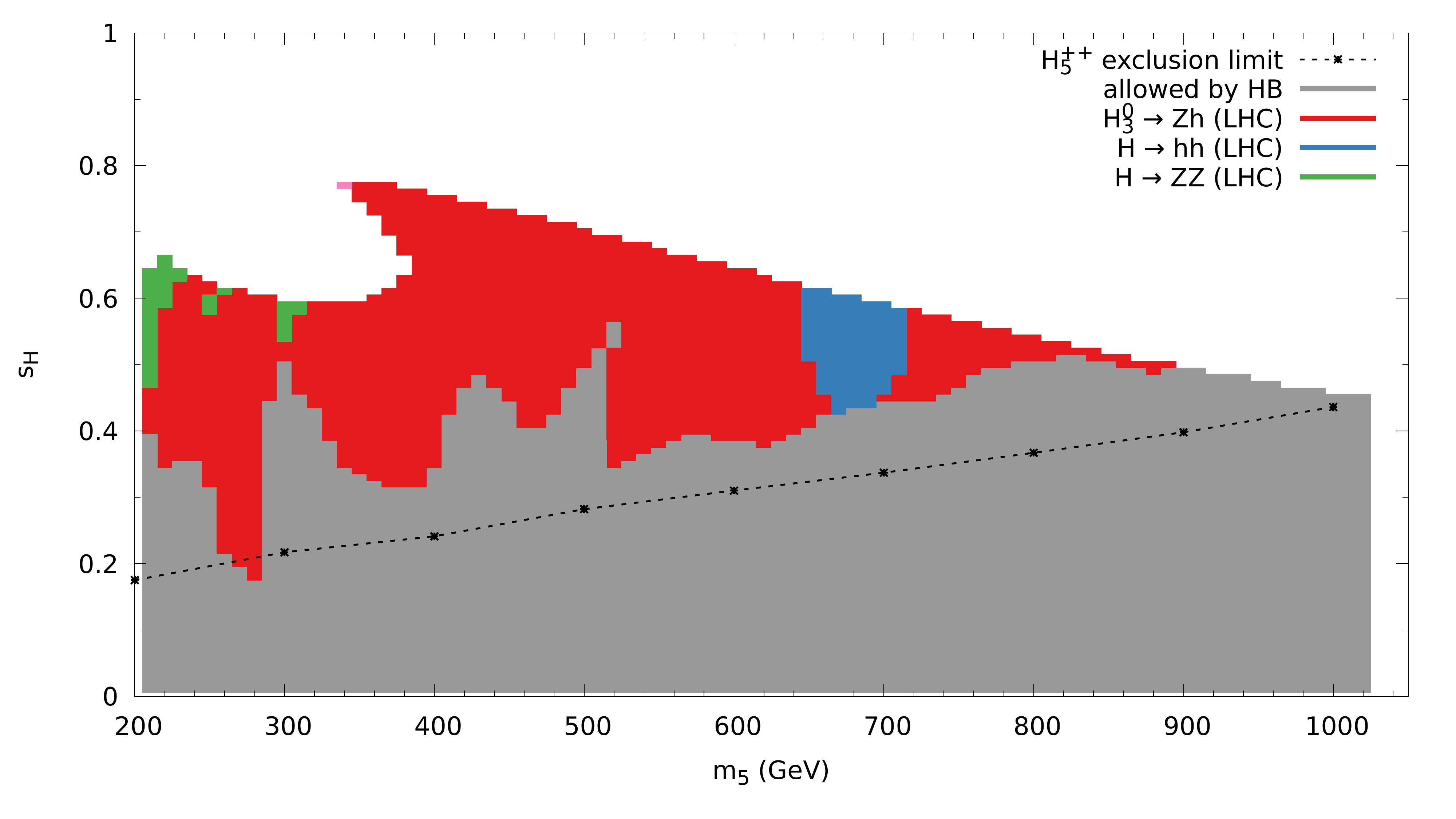}}%
\resizebox{0.5\textwidth}{!}{\includegraphics{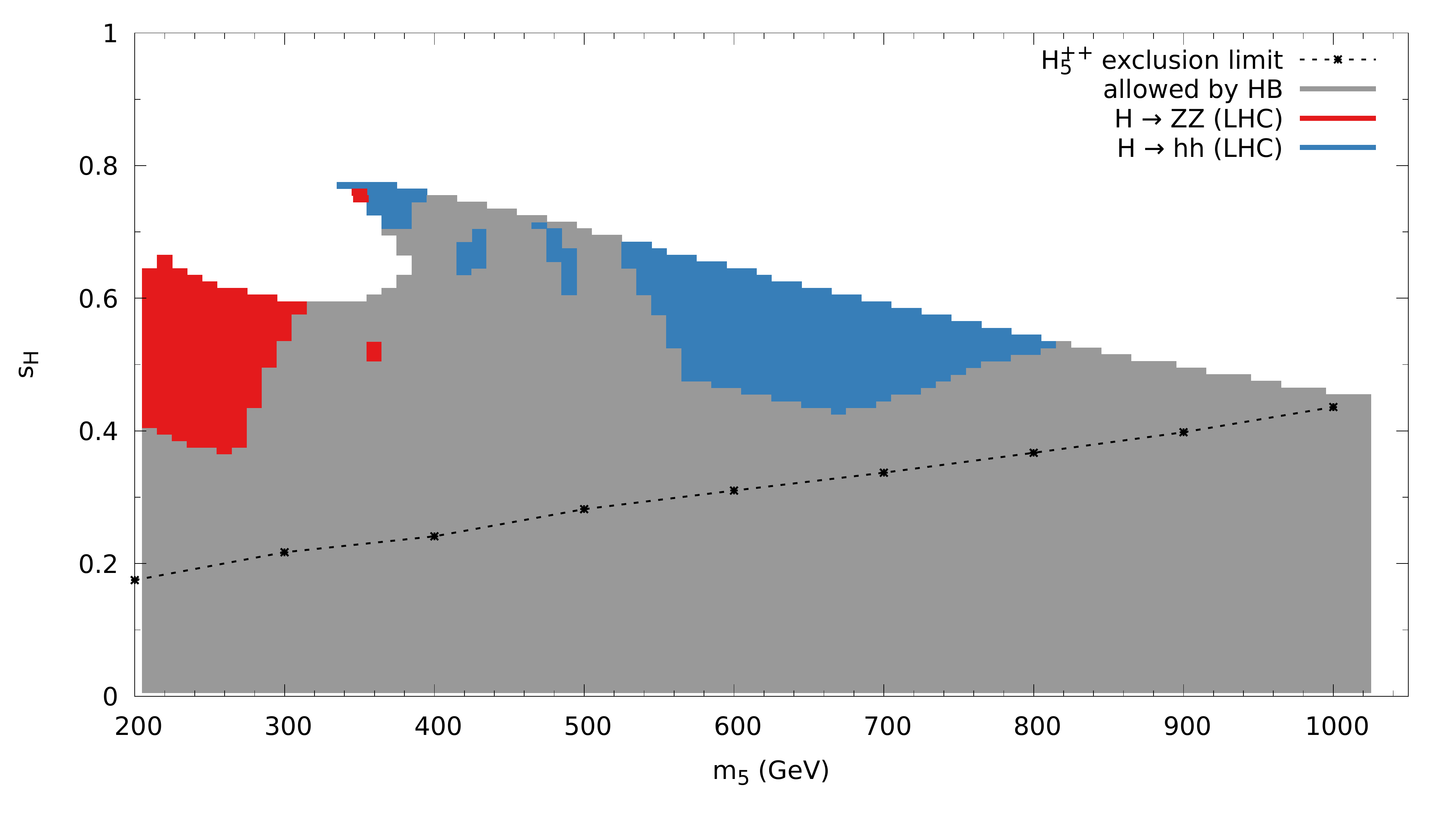}}
    \caption{Left: Excluded parameter regions (in color) from applying HiggsBounds~5.2.0 to all Higgs bosons except the 125~GeV Higgs in the H5plane benchmark.  The region above the black line is excluded by the CMS constraint on VBF $H_5^{\pm \pm} \rightarrow W^{\pm} W^{\pm}$ described in~\autoref{sec:constraints}. The exclusions come from the searches in Refs.~\cite{Khachatryan:2015lba,Aad:2015wra,ATLAS:2016loc,CMS:2017vpy,ATLAS:2016ixk,Khachatryan:2016sey}.
    Right: The same as the left panel but applying HiggsBounds to $H$ alone. The exclusions are from Refs.~\cite{CMS:2017vpy,ATLAS:2016ixk,Aad:2015kna,Khachatryan:2015yea,Khachatryan:2016sey,ATLAS:2016oum}.}
\label{fig:h5plane_1}
\end{figure}

\begin{figure}[!tbp]
\resizebox{0.5\textwidth}{!}{\includegraphics{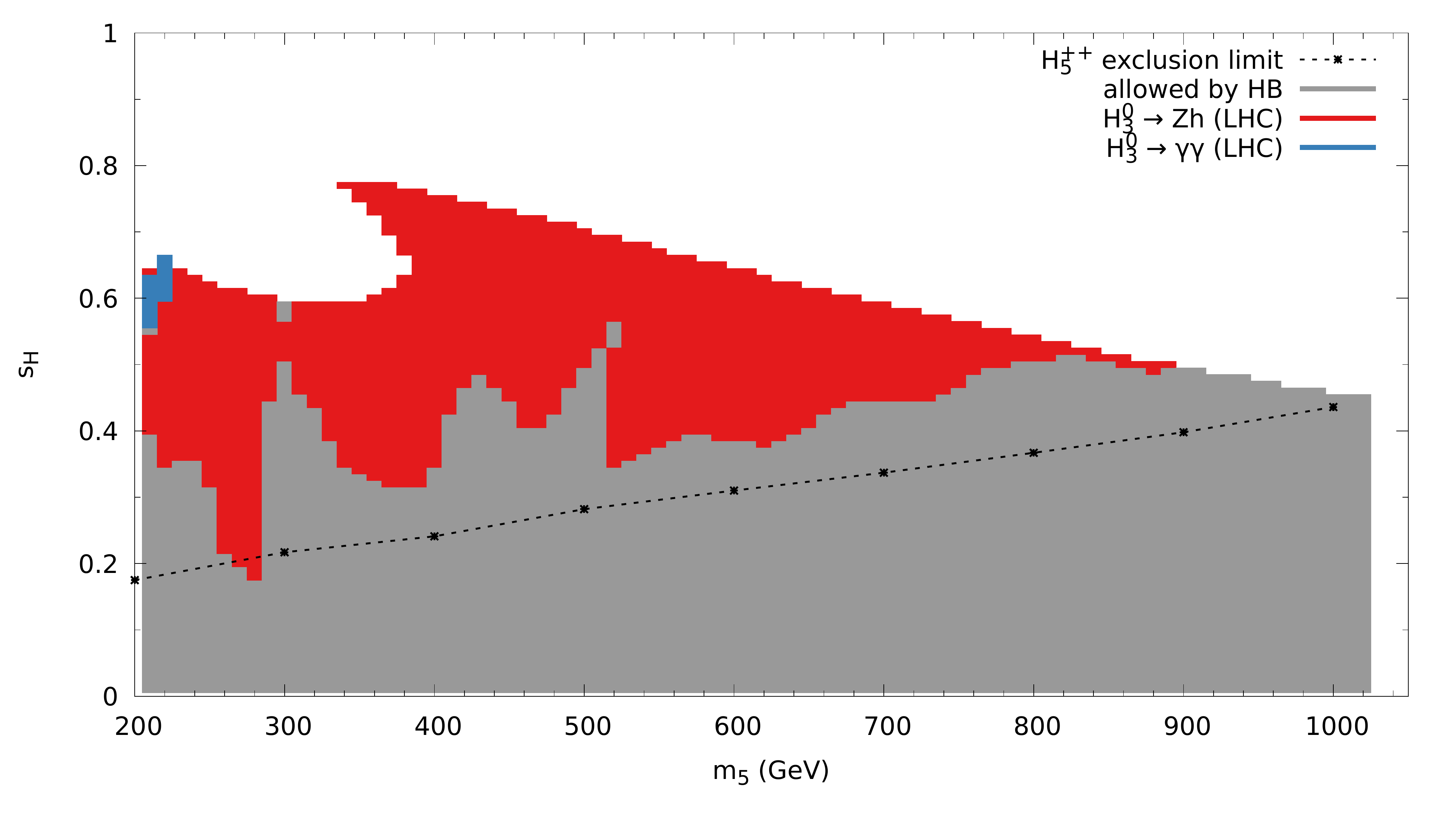}}%
\resizebox{0.5\textwidth}{!}{\includegraphics{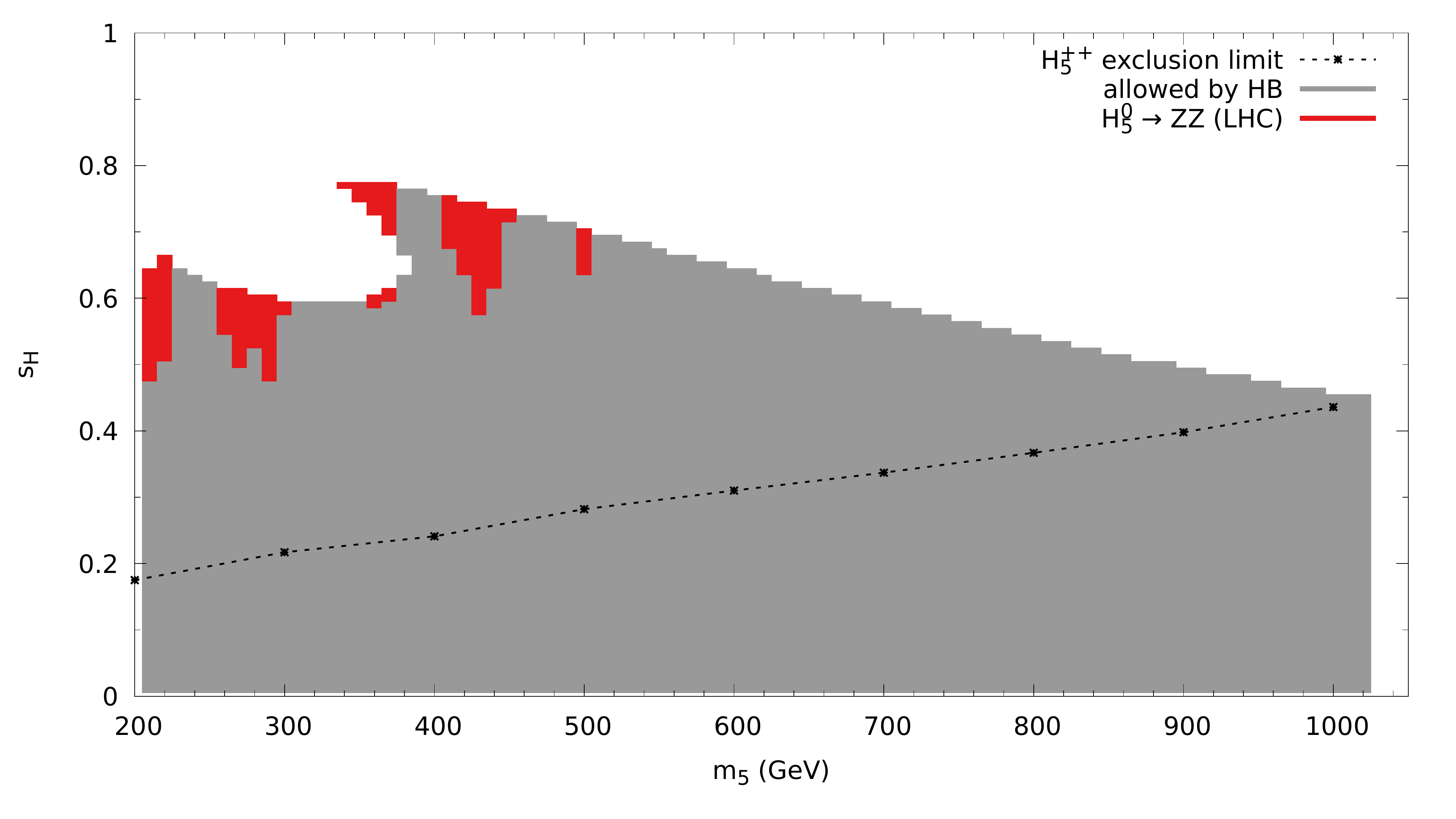}}
    \caption{As in the left panel of \autoref{fig:h5plane_1} but applying HiggsBounds to $H_3^0$ (left) and $H_5^0$ (right) alone. The exclusions are from Refs.~\cite{Khachatryan:2015lba,Aad:2015wra,ATLAS:2016loc,ATLAS:2016eeo} for $H_3^0$ and from Ref.~\cite{ATLAS:2016oum} for $H_5^0$.}
\label{fig:h5plane_2}
\end{figure}

In the left panel of~\autoref{fig:h5plane_1} we show the parameter space in the H5plane benchmark excluded by HiggsBounds 5.2.0 (HB) applied to all Higgs bosons except the 125~GeV Higgs. Exclusion plots for the individual Higgs bosons $H$, $H_3^0$, and $H_5^0$ are also provided in~\autoref{fig:h5plane_1} (right) and \autoref{fig:h5plane_2}; HiggsBounds provides no additional exclusions for $H_3^{\pm}$, $H_5^{\pm}$, or $H_5^{\pm\pm}$ in this benchmark. Excluded regions of the parameter space are color-coded to indicate the most sensitive excluding search channel as reported by HiggsBounds. The exclusion curve from the CMS constraint on VBF production of $H_5^{\pm \pm}$ decaying to $W^{\pm} W^{\pm}$ (discussed in~\autoref{sec:constraints}) is also plotted. Throughout most of the parameter space of this benchmark, HiggsBounds does not exclude any area that is not already excluded by this constraint; the one exception is a small region around $m_5 = 280$ GeV and $s_H = 0.2$ that is excluded by $H_3^0 \rightarrow Z h$. The right panel of~\autoref{fig:h5plane_1} indicates that $H \rightarrow hh$ is a potentially important search channel (excluded region shown in blue), but not as powerful as the $H_5^{\pm\pm}$ constraint in this benchmark.

\subsection{Low $m_5$ benchmark}
\begin{figure}[!tbp]
\includegraphics[width=0.8\textwidth]{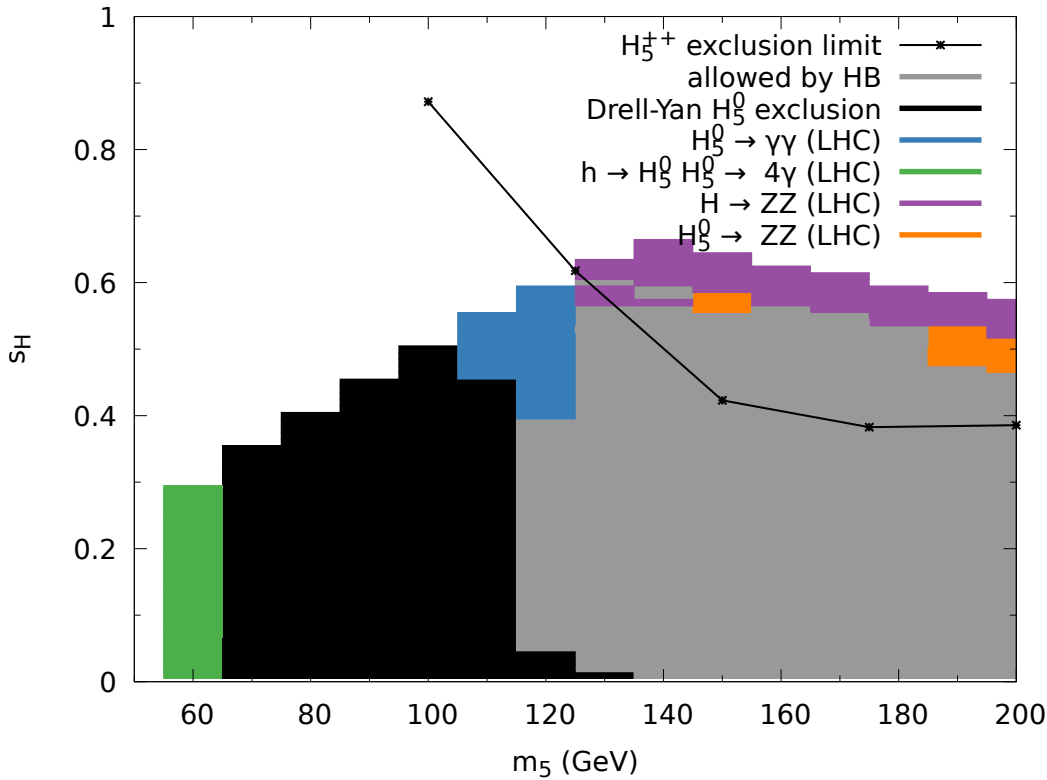}
    \caption{Excluded parameter regions (in color) from applying HiggsBounds~5.2.0 to all Higgs bosons except the 125~GeV Higgs in the low-$m_5$ benchmark. The region above the black line is excluded by the ATLAS constraint on VBF $H_5^{\pm \pm} \rightarrow W^{\pm} W^{\pm}$ described in~\autoref{sec:constraints}. The region excluded by Drell-Yan production of $H_5^0$ decaying to diphotons (see~\autoref{sec:constraints}) is shown in black. The HiggsBounds exclusions come from the searches in Refs.~\cite{ALEPH:2002gcw,CMS:2015ocq,CMS:2016ilx,Aad:2014ioa,Aad:2015kna,ATLAS:2016oum,Aad:2015bua,Aaboud:2017rel}.}
\label{fig:lowplane_1}
\end{figure}

\begin{figure}[!tbp]
\resizebox{0.5\textwidth}{!}{\includegraphics{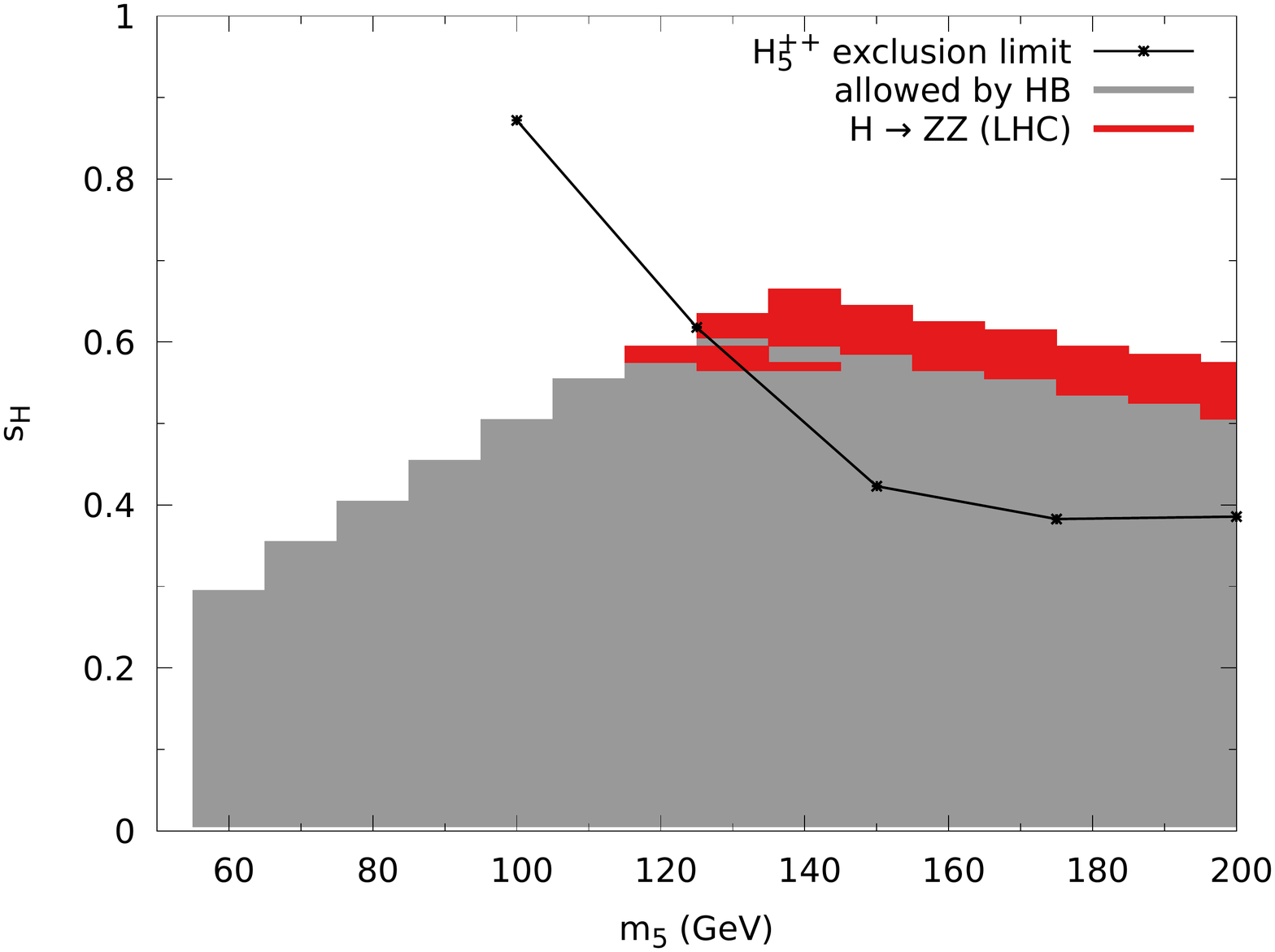}}%
\resizebox{0.5\textwidth}{!}{\includegraphics{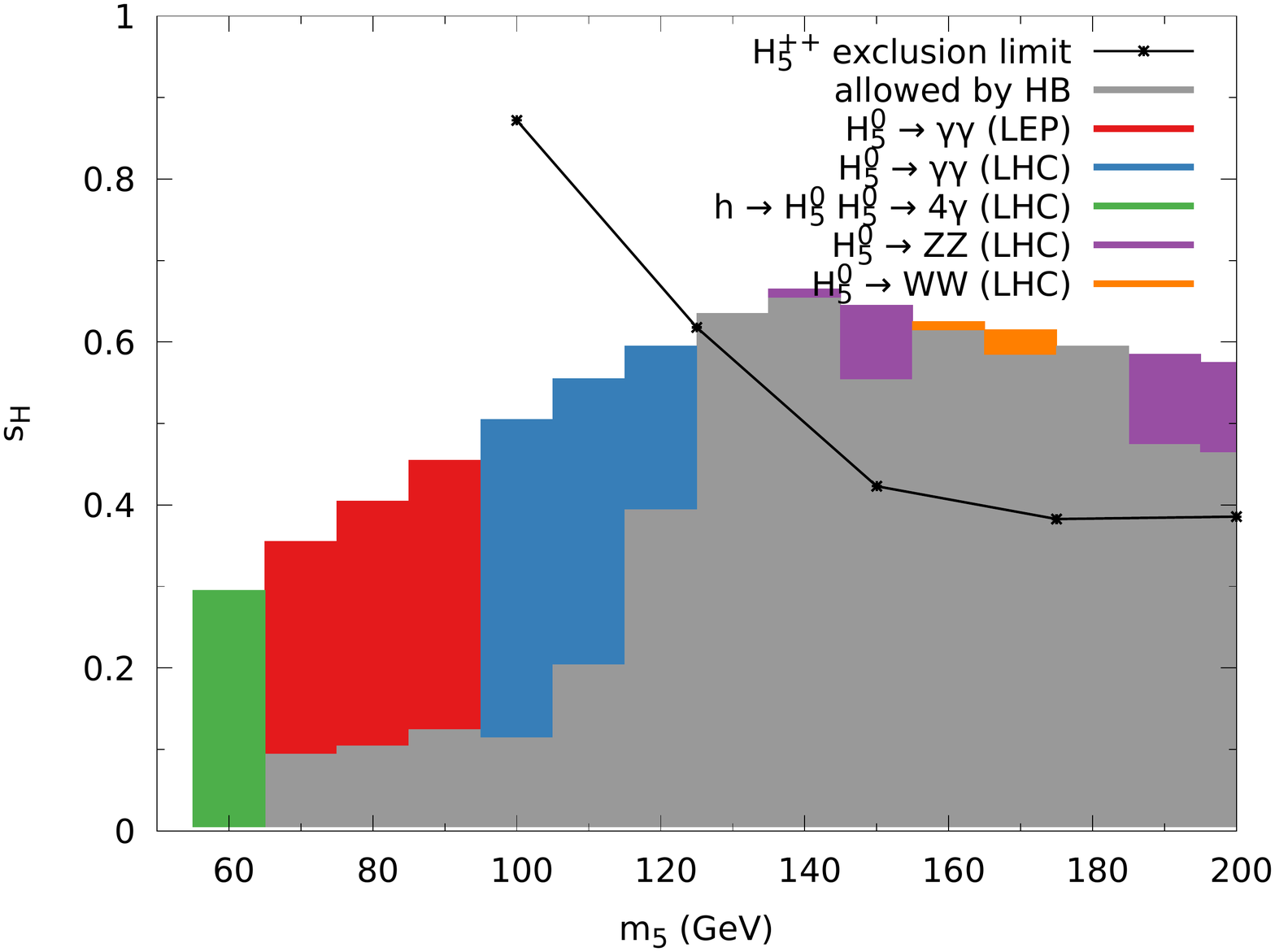}}
    \caption{As in \autoref{fig:lowplane_1} but applying HiggsBounds to $H$ (left) and $H_5^0$ (right) alone.  The exclusions are from Refs.~\cite{Aad:2015kna,Aaboud:2017rel} for $H$ and from Refs.~\cite{CMS:2016ilx,CMS:2013yea,ALEPH:2002gcw,CMS:2015ocq,Aad:2014ioa,ATLAS:2016oum,Aad:2015bua} for $H_5^0$.}
\label{fig:lowplane_2}
\end{figure}

In \autoref{fig:lowplane_1} we show the parameter space in the low-$m_5$ benchmark excluded by HiggsBounds 5.2.0 applied to all Higgs bosons except the 125~GeV Higgs.  Excluded regions are again color-coded to indicate the most sensitive excluding search channel as reported by HiggsBounds.  Also shown are the exclusion from the ATLAS constraint on VBF $H_5^{\pm \pm} \rightarrow W^{\pm} W^{\pm}$ described in~\autoref{sec:constraints} (black curve) and the region excluded by Drell-Yan production of $H_5^0$ decaying to diphotons described in~\autoref{sec:constraints} (solid black region).  Most of the parameter space of the low-$m_5$ benchmark with $m_5 \lesssim 120$~GeV is excluded by the latter constraint. Exclusion plots for the individual Higgs bosons $H$ and $H_5^0$ from HiggsBounds are also provided in~\autoref{fig:lowplane_2}; HiggsBounds provides no additional exclusions for $H_3^0$, $H_3^{\pm}$, $H_5^{\pm}$, or $H_5^{\pm\pm}$ in this benchmark.  The most interesting HiggsBounds constraints are from LHC searches for $h \to H_5^0H_5^0 \to 4\gamma$ (green) and $H_5^0 \to \gamma\gamma$ with $H_5^0$ produced singly (blue).  

\subsection{Full scan, high mass}
\begin{figure}[!tbp]
\includegraphics[width=0.8\textwidth]{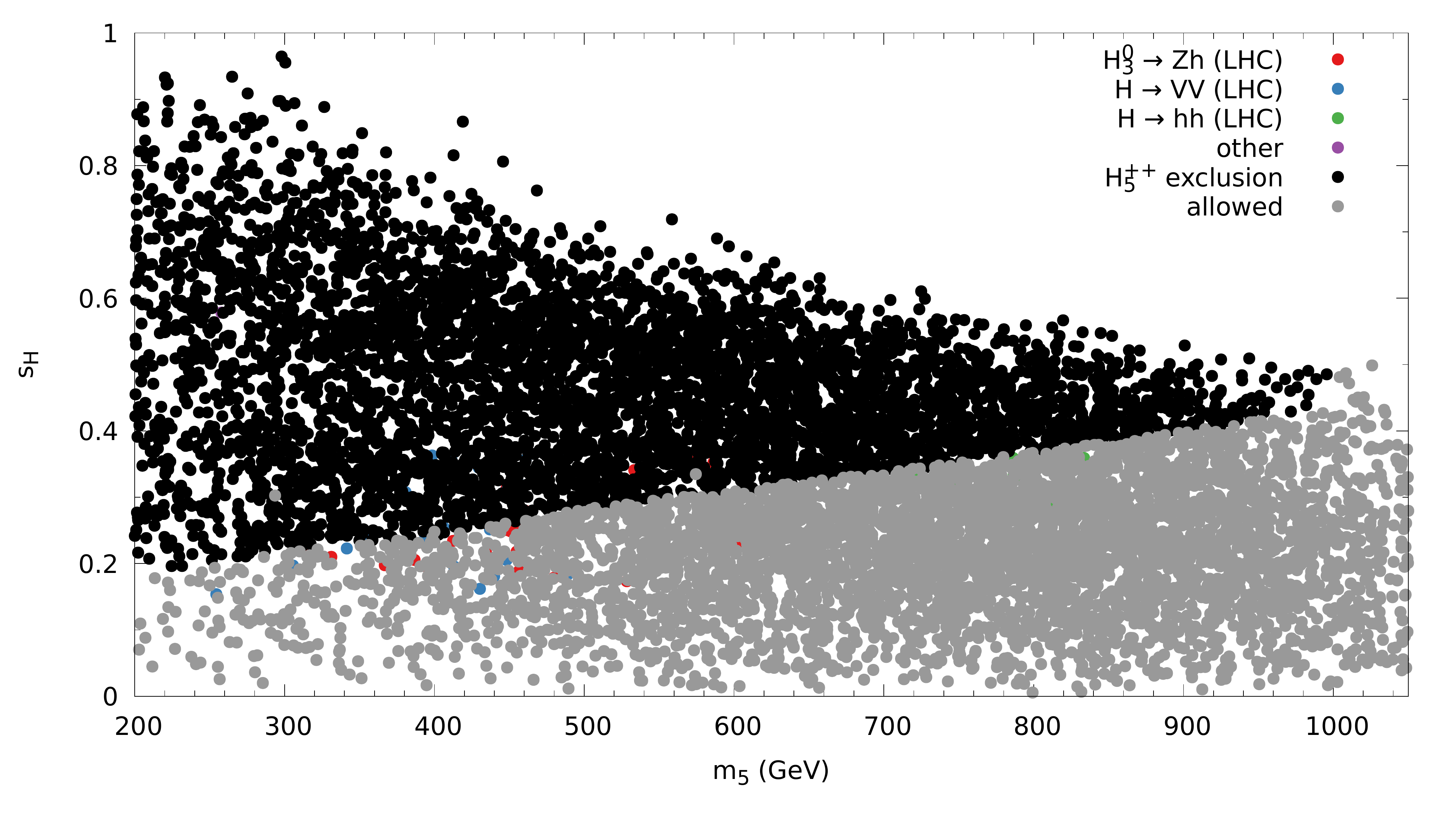}
    \caption{Allowed parameter points (gray, plotted on top) from a general scan of 10,000 points with $m_5 \in [200, 1050)$~GeV in the GM model.  Points excluded by the CMS constraint on VBF production of $H_5^{\pm \pm}$ decaying to $W^{\pm} W^{\pm}$~\cite{Sirunyan:2017ret} (described in~\autoref{sec:constraints}) are shown in black.  Points allowed by this constraint but excluded by HiggsBounds are shown in color.  The HiggsBounds exclusions come from the searches in Refs.~\cite{Khachatryan:2015lba,Aad:2015wra,CMS:2017vpy,Aaboud:2017rel,Aad:2015kna,Khachatryan:2015cwa,ATLAS:2016ixk,Aaboud:2018gjj}.}
\label{fig:highmass_1}
\end{figure}

\begin{figure}[!tbp]
\resizebox{0.5\textwidth}{!}{\includegraphics{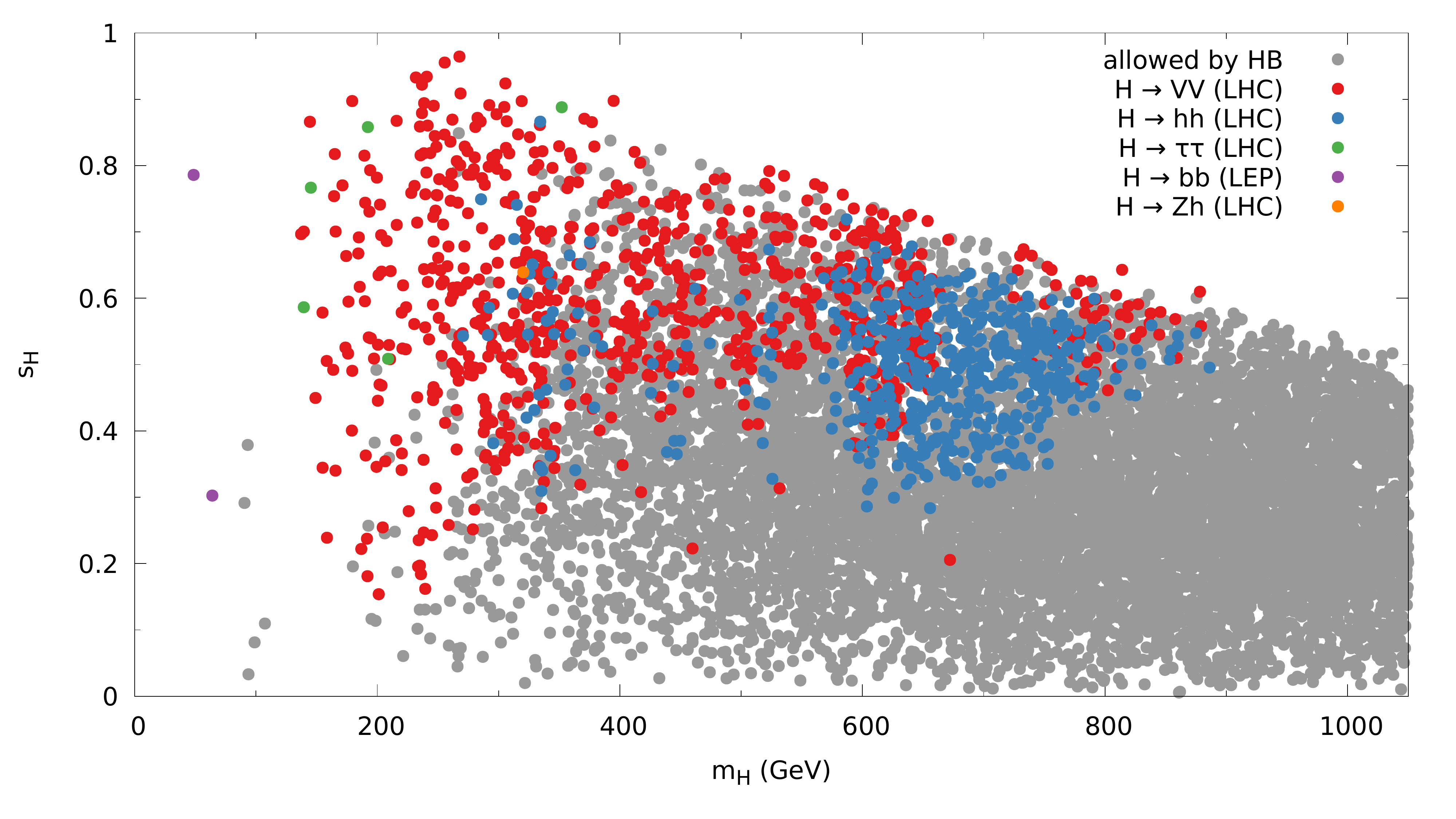}}%
\resizebox{0.5\textwidth}{!}{\includegraphics{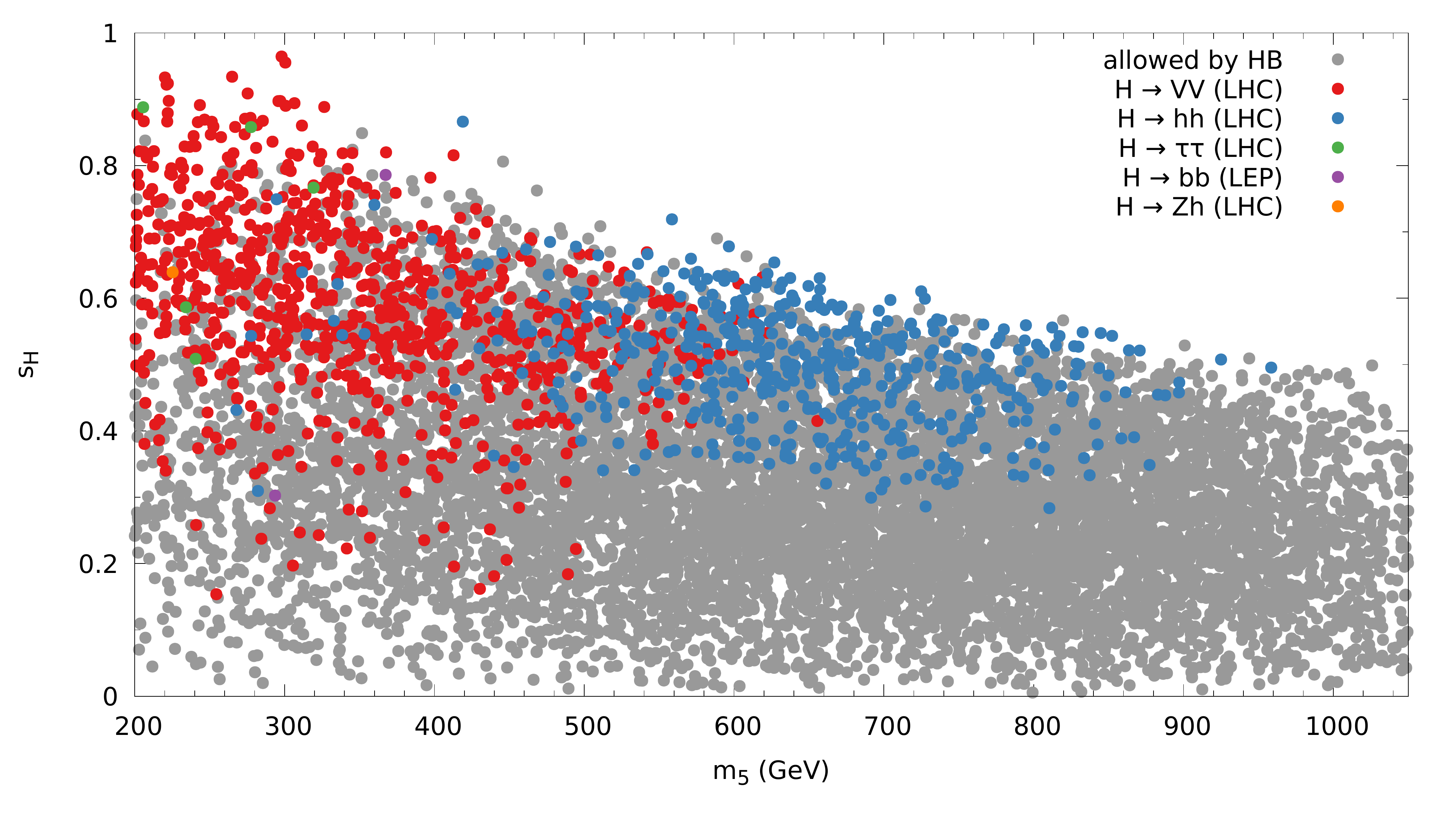}}
    \caption{Excluded parameter points (in color, plotted on top) from applying HiggsBounds~5.2.0 to $H$ alone in a general scan of 10,000 points in the GM model with $m_5 \in [200, 1050)$ GeV, plotted as a function of $m_H$ (left) and $m_5$ (right). The exclusions come from Refs.~\cite{CMS:2017vpy,Aaboud:2017rel,ATLAS:2016oum,Aad:2015kna,Khachatryan:2015cwa,CMS:2013yea,ATLAS:2016ixk,Khachatryan:2015yea,Khachatryan:2016sey,ATLAS:2016qmt,CMS:2015mca,Schael:2006cr,Khachatryan:2015lba}.}
\label{fig:highmass_2}
\end{figure}

\begin{figure}
\resizebox{0.5\textwidth}{!}{\includegraphics{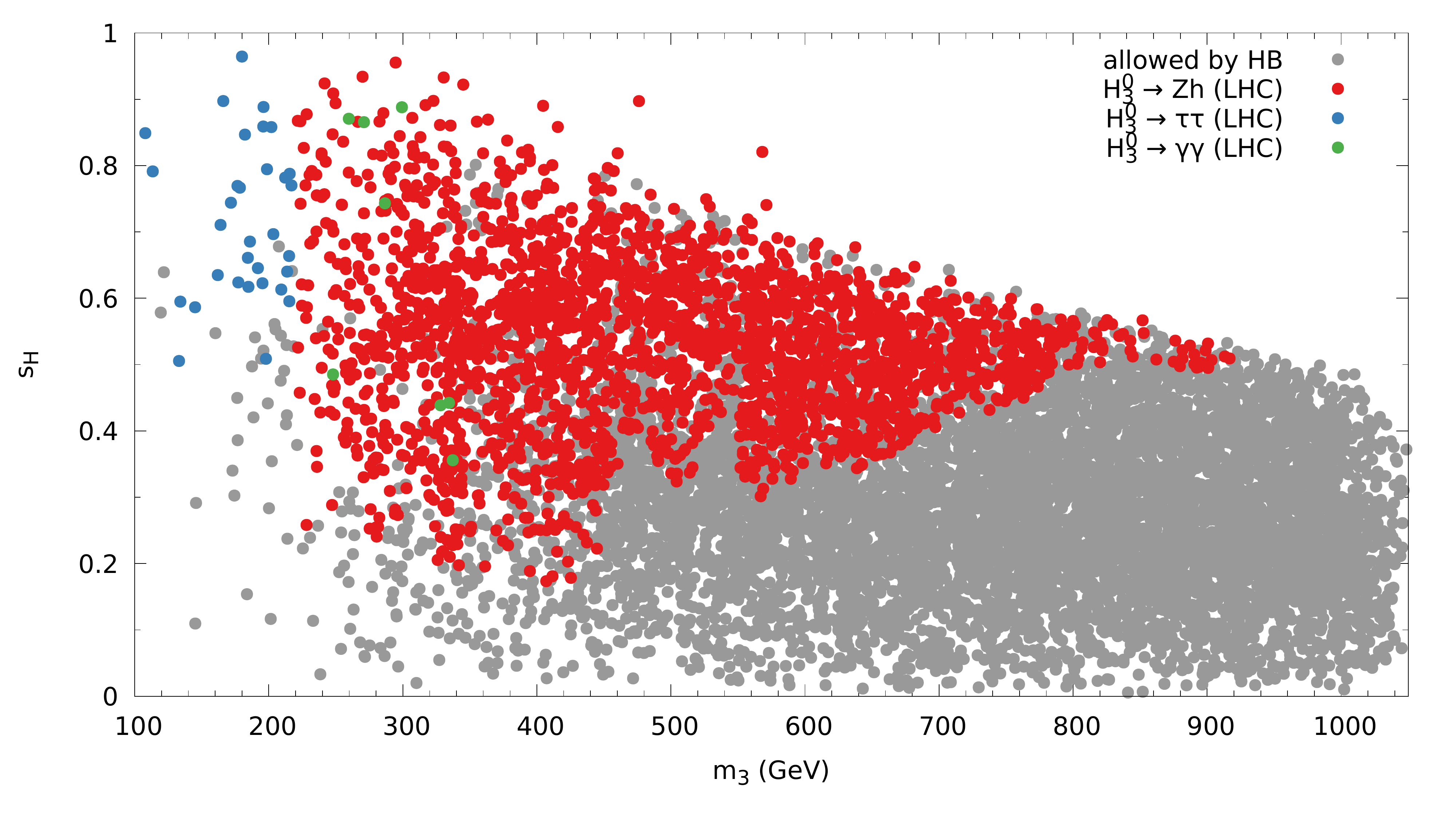}}%
\resizebox{0.5\textwidth}{!}{\includegraphics{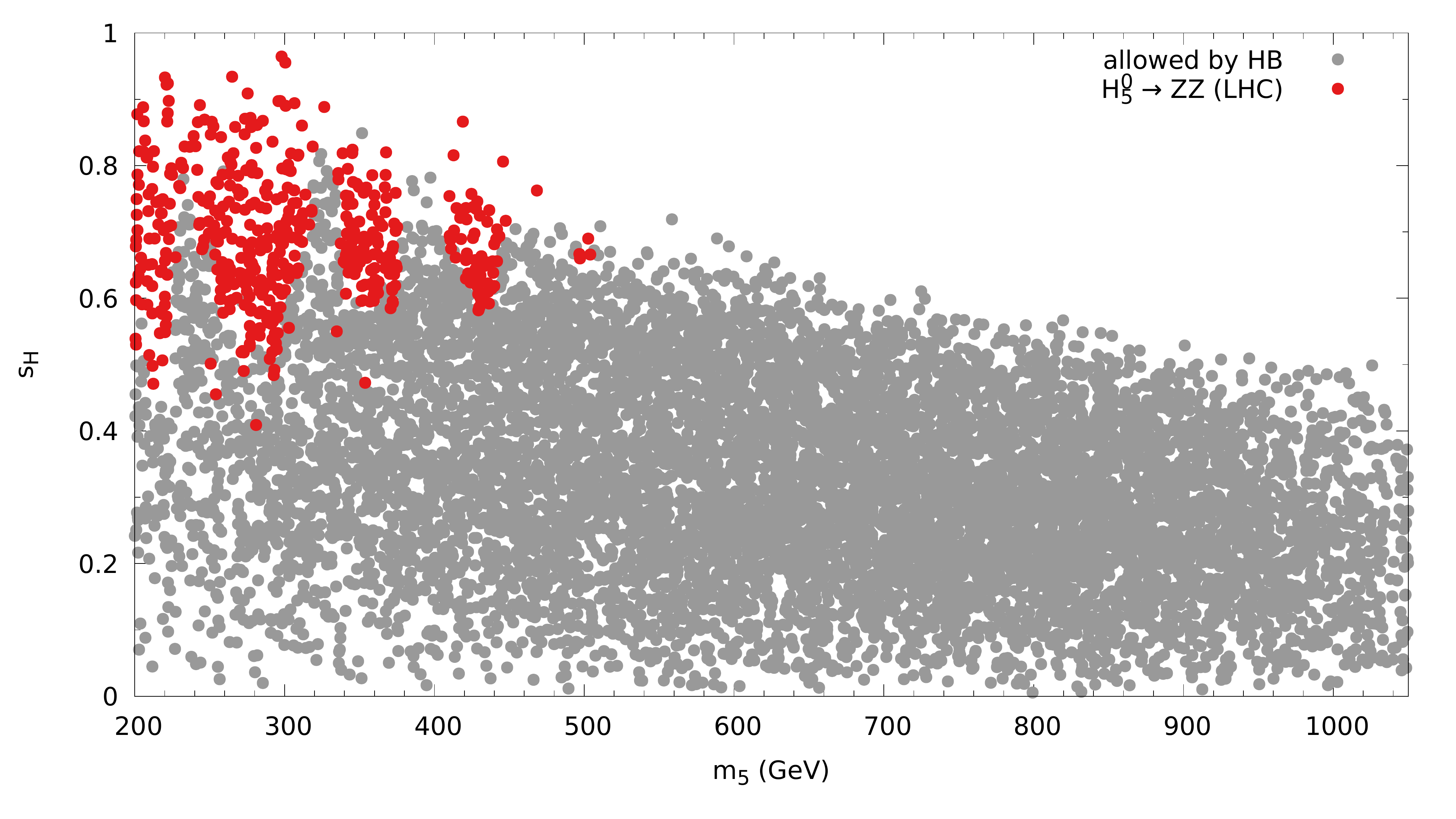}}
    \caption{Excluded parameter points (in color, plotted on top) from applying HiggsBounds~5.2.0 to $H_3^0$ (left) or $H_5^0$ (right) alone in a general scan of 10,000 points in the GM model with $m_5 \in [200, 1050)$ GeV.  The exclusions come from Refs.~\cite{Aad:2015wra,ATLAS:2016loc,Khachatryan:2015lba,Khachatryan:2016are,CMS:2015mca,Aad:2014vgg,ATLAS:2016eeo,Khachatryan:2015qba} for $H_3^0$ and from Ref.~\cite{ATLAS:2016oum} for $H_5^0$.}
\label{fig:highmass_3}
\end{figure}

To test the generality of the results in the benchmarks, we performed two general scans over the full 7-dimensional parameter space of the GM model. In the first scan, 10,000 points that satisfied all theoretical constraints were randomly generated in the region $m_5 \in [200, 1050)$ GeV, the same $m_5$ range as in the H5plane benchmark. \autoref{fig:highmass_1} shows the allowed (gray) and excluded points.  Unlike in the H5plane benchmark, the VBF $H_5^{\pm \pm} \rightarrow W^{\pm} W^{\pm}$ constraint from CMS cannot be represented as a curve, since the points do not represent a two-dimensional slice of the parameter space; instead, points excluded by this constraint are shown in black. The points excluded by this constraint nevertheless very nearly coincide with the exclusion curve in the H5plane benchmark plots. A handful of points allowed by the $H_5^{\pm\pm}$ constraint are excluded by other processes, including $H_3^0 \rightarrow Zh$ (red) and $H \rightarrow hh$ (green). A considerable region of the parameter space remains unexcluded.  Exclusion plots obtained by applying HiggsBounds to $H$, $H_3^0$, or $H_5^0$ alone are shown in \autoref{fig:highmass_2} and \autoref{fig:highmass_3}; the patterns of exclusions are very similar to the corresponding plots in the H5plane benchmark.

\subsection{Full scan, low mass}
\begin{figure}[!tbp]
\includegraphics[width=0.8\textwidth]{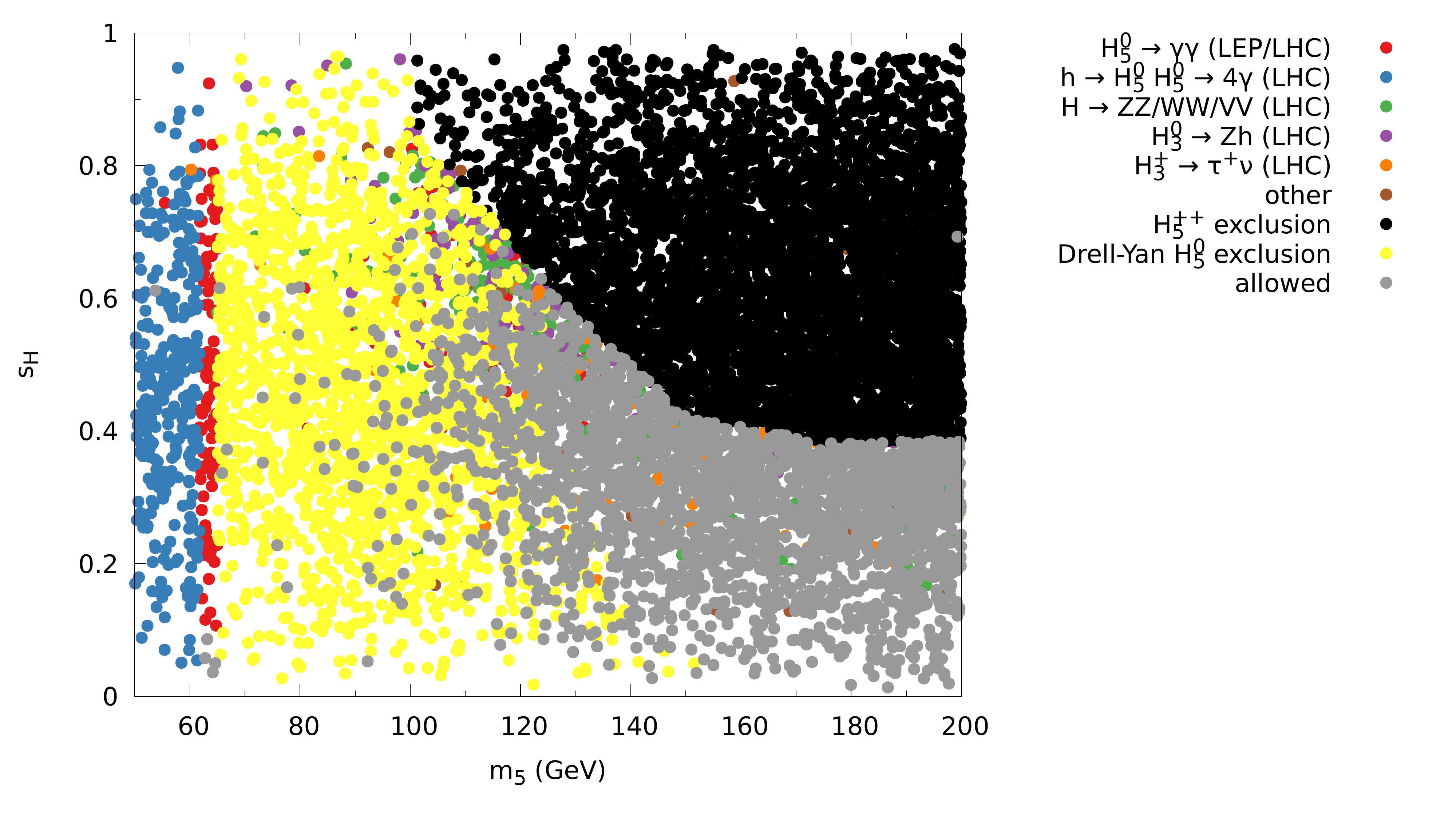}
    \caption{Allowed parameter points (gray, plotted on top) from a general scan of 10,000 points with $m_5 \in [50, 200)$~GeV in the GM model. Points excluded by the ATLAS constraint on VBF $H_5^{\pm \pm}$ decaying to $W^{\pm} W^{\pm}$~\cite{Chiang:2014bia,Aad:2014zda} are shown in black and points excluded by the constraint on Drell-Yan production of $H_5^0$ decaying to diphotons (see~\autoref{sec:constraints}) are shown in yellow. The rest of the points shown in color are excluded by HiggsBounds 5.2.0 and come from Refs.~\cite{ALEPH:2002gcw,CMS:2015ocq,Aad:2014ioa,Aad:2015bua,ATLAS:2016oum,CMS:2017vpy,Aad:2015kna,Aaboud:2017rel,Khachatryan:2015cwa,CMS:2013yea,Khachatryan:2015lba,Aad:2015wra,Aaboud:2018gjj,CMS:2014cdp,CMS:2015mca,Aad:2012tfa,Chatrchyan:2012xdj,Schael:2006cr,ATLAS:2016eeo,Khachatryan:2016sey,Khachatryan:2015qba,CMS:aya}.}
\label{fig:lowmass_1}
\end{figure}

For the second general scan, 10,000 points that satisfied all theoretical constraints were randomly generated in the region $m_5 \in [50, 200)$ GeV, the same $m_5$ range as in the low-$m_5$ benchmark. \autoref{fig:lowmass_1} shows the allowed (gray) and excluded points.  Points excluded by the ATLAS constraint on VBF $H_5^{\pm \pm} \rightarrow W^{\pm} W^{\pm}$ are shown in black. Again, the points excluded by this constraint very nearly coincide with the exclusion curve in the low-$m_5$ benchmark plots.  In contrast to the low-$m_5$ benchmark, there are many allowed points with $m_5 < 120$ GeV, some with $s_H$ as high as $0.7$.  These points appear when accidental cancellations suppress the branching ratio of $H_5^0 \to \gamma\gamma$, allowing them to evade the constraint from Drell-Yan $H_5^0 \to \gamma\gamma$. Another difference from the benchmark is the presence of several points excluded by HiggsBounds from the decay of a charged Higgs, $H_3^{+} \rightarrow \tau^{+} \nu$. 

\begin{figure}[!tbp]
\resizebox{0.5\textwidth}{!}{\includegraphics{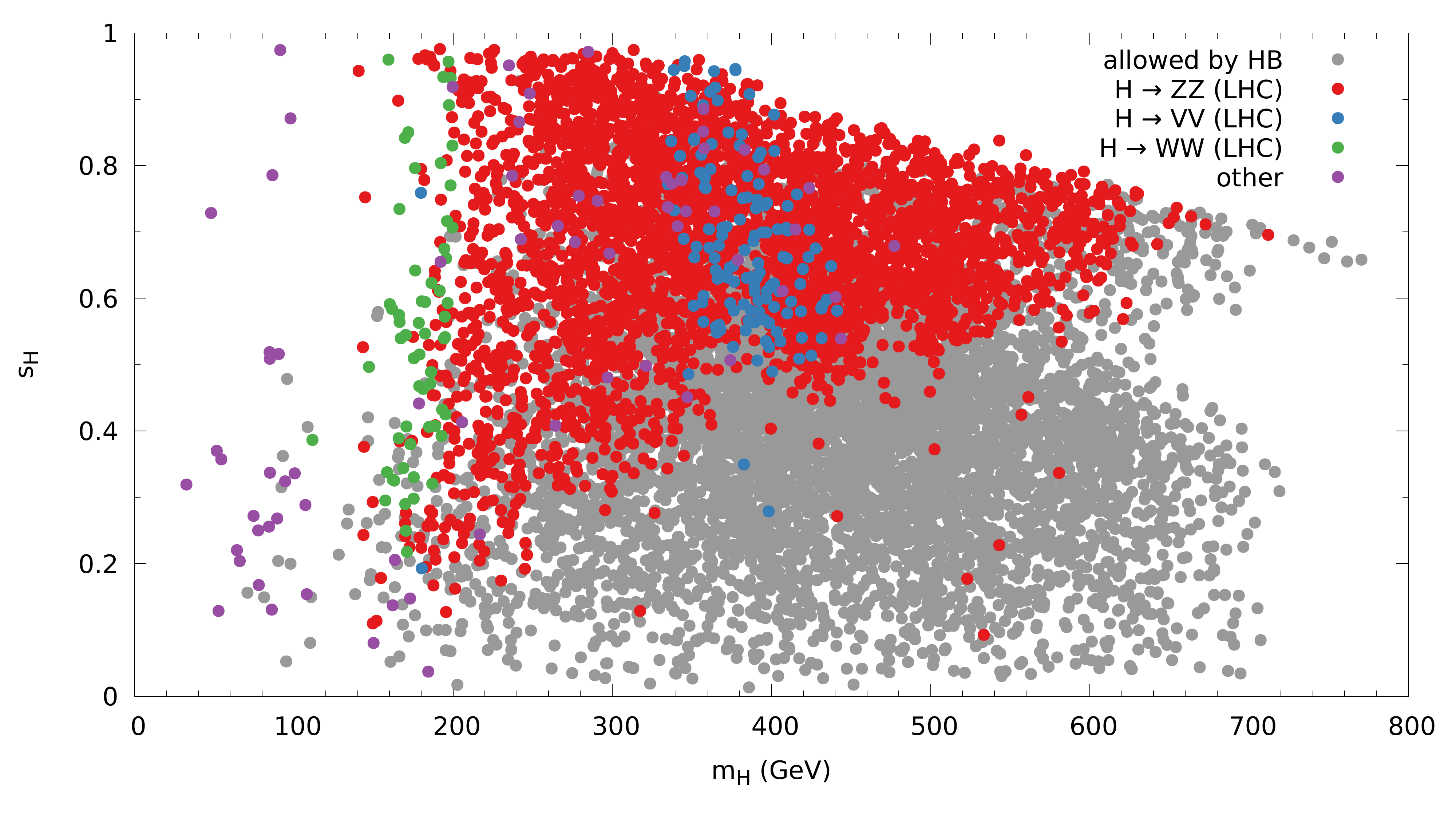}}%
\resizebox{0.5\textwidth}{!}{\includegraphics{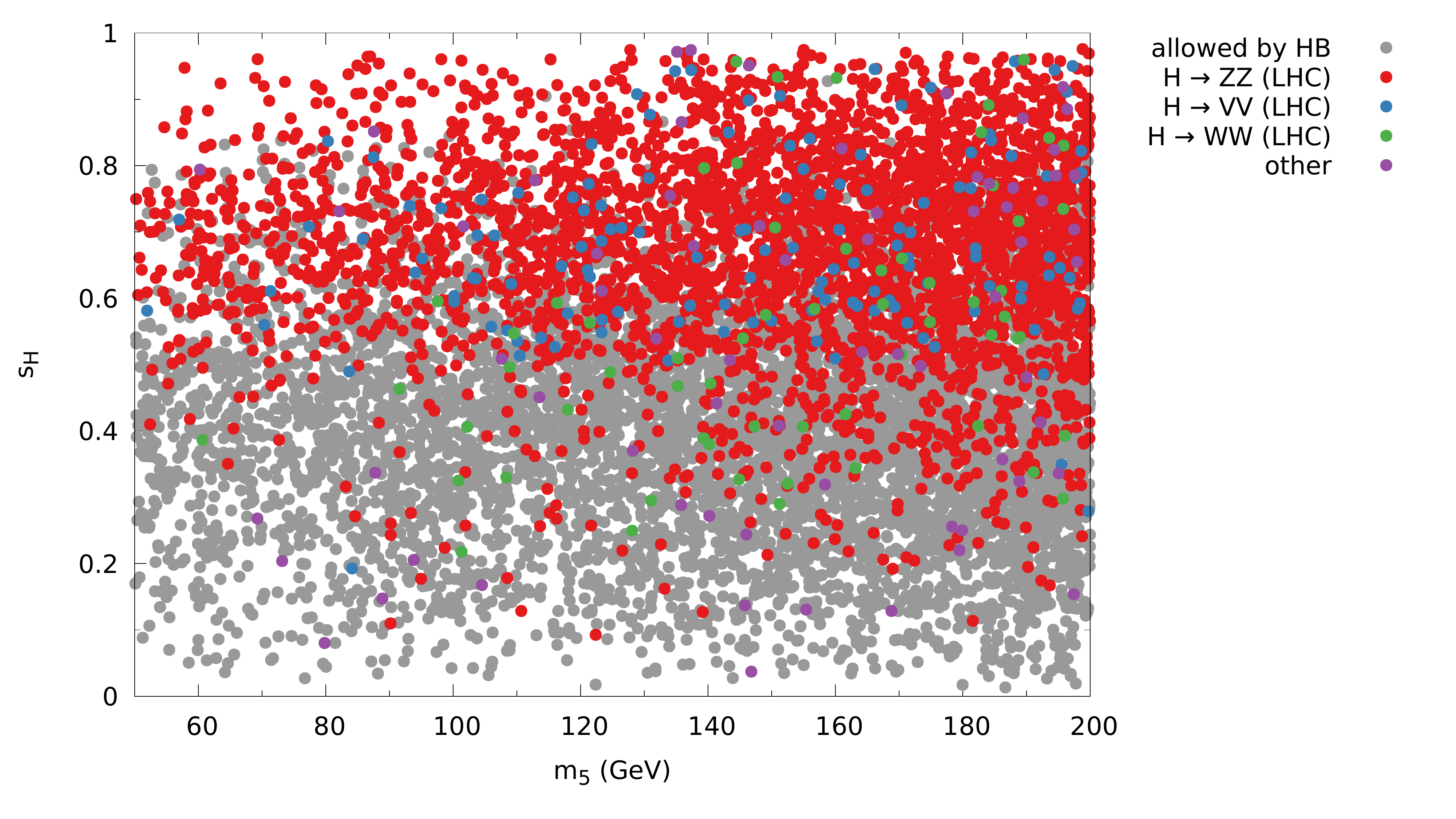}}
    \caption{Excluded parameter points (in color, plotted on top) from applying HiggsBounds 5.2.0 to $H$ alone in a general scan of 10,000 points in the GM model with $m_5 \in [50, 200)$ GeV, plotted as a function of $m_H$ (left) and $m_5$ (right). The exclusions come from searches reported in Refs.~\cite{ATLAS:2016oum,CMS:2017vpy,Aad:2015kna,Aaboud:2017rel,Khachatryan:2015cwa,CMS:2013yea,Khachatryan:2016are,Khachatryan:2016sey,CMS:2015mca,Schael:2006cr,Khachatryan:2015lba,ATLAS:2016qmt,ATLAS:2016eeo,Khachatryan:2015qba,CMS:2015ocq,CMS:aya,ALEPH:2002gcw}.}
\label{fig:lowmass_2}
\end{figure}

\begin{figure}[!tbp]
\resizebox{0.5\textwidth}{!}{\includegraphics{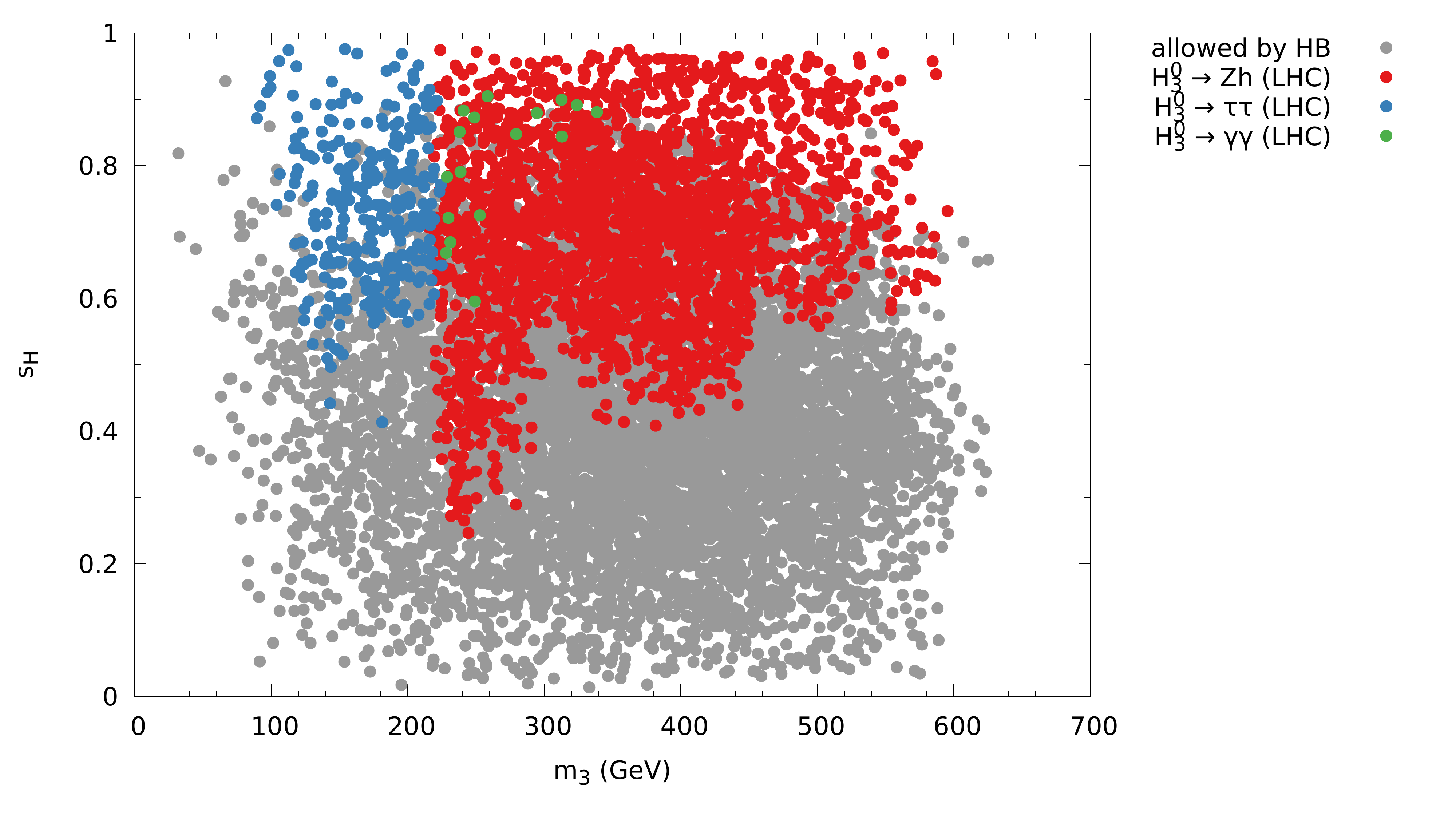}}%
\resizebox{0.5\textwidth}{!}{\includegraphics{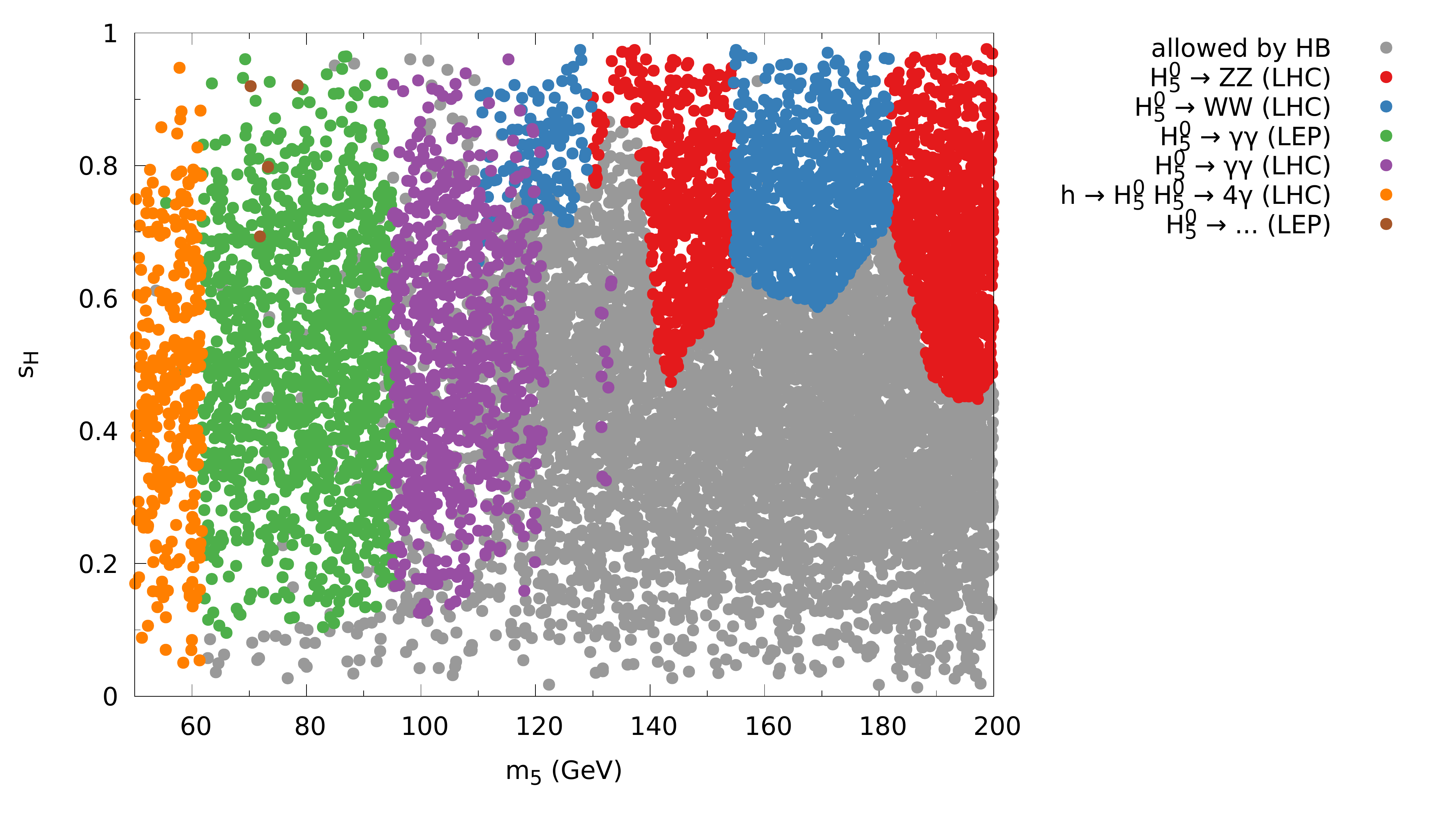}}
    \caption{Excluded parameter points (in color, plotted on top) from applying HiggsBounds 5.2.0 to $H_3^0$ (left) or $H_5^0$ (right) alone in a general scan of 10,000 points in the GM model with $m_5 \in [50, 200)$ GeV.  The exclusions come from Refs.~\cite{Khachatryan:2015lba,Aad:2015wra,Khachatryan:2016are,CMS:2015mca,CMS:2017epy,Aad:2014vgg,Aad:2014ioa,ATLAS:2016eeo} for $H_3^0$ and from Refs.~\cite{CMS:2016ilx,ATLAS:2016oum,CMS:2013yea,ALEPH:2002gcw,CMS:2015ocq,Aad:2014ioa,Aad:2015bua,Abbiendi:2002qp} for $H_5^0$.}
\label{fig:lowmass_3}
\end{figure}

Exclusion plots obtained by applying HiggsBounds to $H$, $H_3^0$, or $H_5^0$ alone are shown in \autoref{fig:lowmass_2} and \autoref{fig:lowmass_3}.  Points excluded by searches for $H$ or $H_5^0$ alone come from the same search channels as in the low-$m_5$ benchmark.  However, while no points in the low-$m_5$ benchmark were exlcuded by HiggsBounds from searches for $H_3^0$, in the general scan HiggsBounds excludes many points, mainly through searches for $H_3^0 \rightarrow Z h$ and $H_3^0 \rightarrow \tau \tau$.

\section{Constraints from HiggsSignals 2.2.1}
\label{sec:HS}

\begin{figure}[!tbp]
\resizebox{0.5\textwidth}{!}{\includegraphics{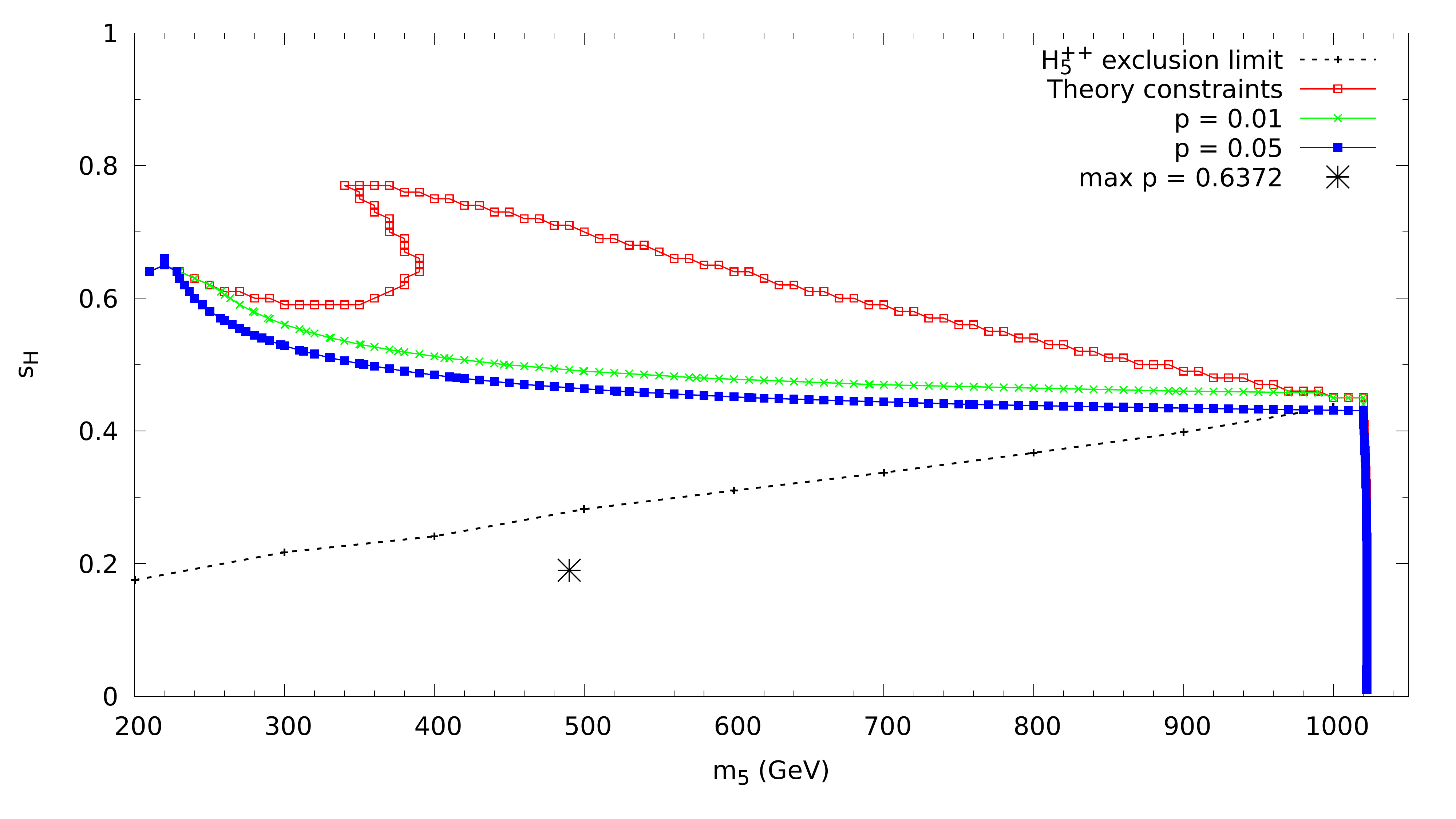}}%
\resizebox{0.5\textwidth}{!}{\includegraphics{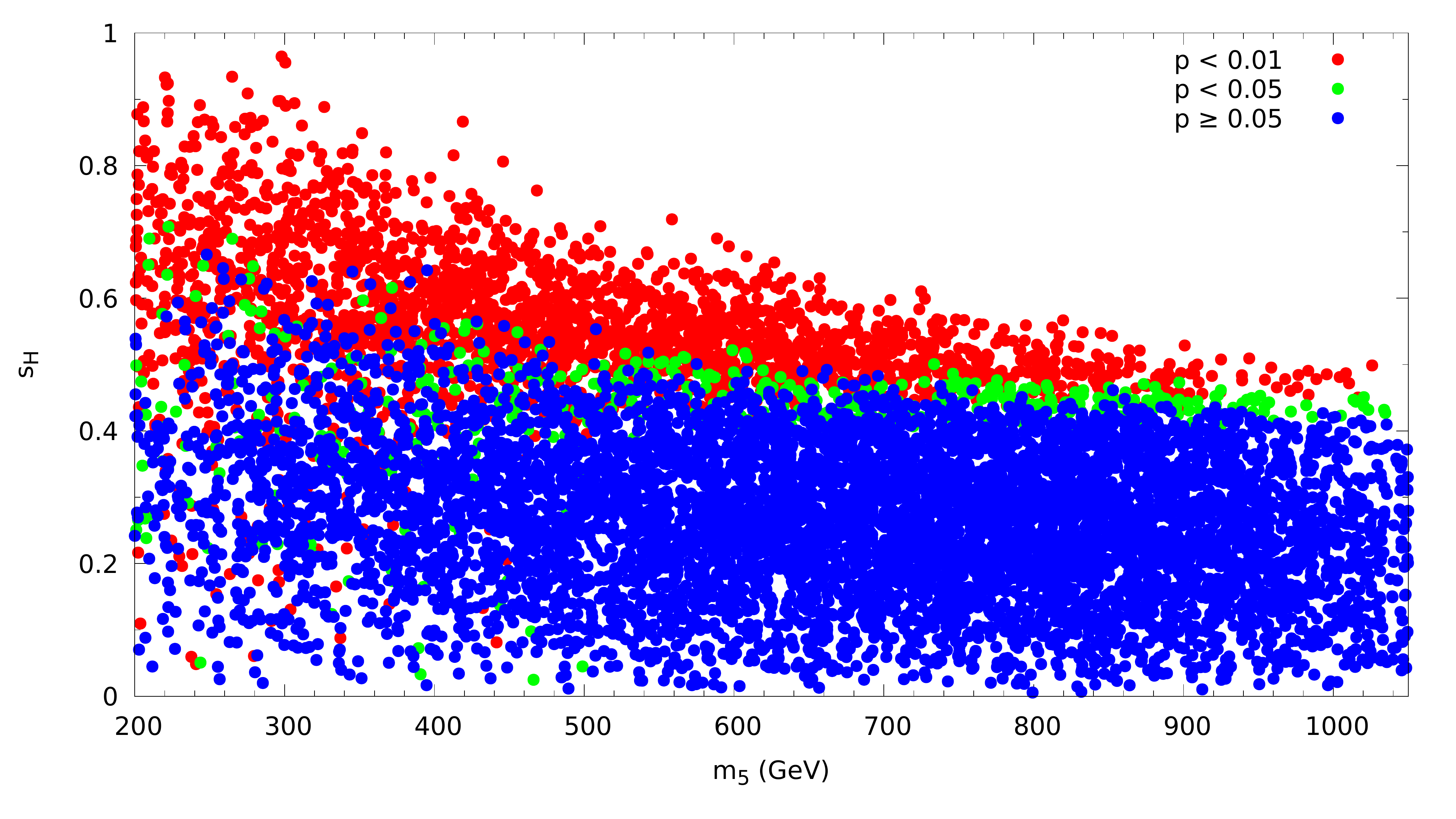}}
\caption{Left: contours of the $p$-value from a fit to experimental measurements of the 125~GeV Higgs boson's couplings in the H5plane benchmark calculated with HiggsSignals. The region above the black dotted line is excluded by a CMS search for VBF production of $H_5^{\pm\pm}$ decaying to $W^{\pm} W^{\pm}$ as described in~\autoref{sec:constraints}. Right: as in the left panel but for a general scan of 10,000 points in the GM model with $m_5 \in [200, 1050)$ GeV.}
\label{fig:hs_high}
\end{figure}

We now apply the measured signal strengths of the 125~GeV Higgs boson using HiggsSignals~2.2.1 to each of our benchmarks and scans.  The constraints on the H5plane benchmark from HiggsSignals are shown in~\autoref{fig:hs_high} (left), where we plot contours of the $p$-value computed by HiggsSignals with two free parameters.  All the parameter space in the H5plane with $s_H \leq 0.4$ is allowed at the 95\% confidence level ($p > 0.05$) by the measured Higgs signal strengths. The maximum $p$-value obtained in the H5plane benchmark is 0.64, a slightly better fit to the data than for the SM Higgs, which yields $p = 0.42$. The picture in the general scan with $m_5 \in [200, 1050)$ GeV is very similar (right panel of \autoref{fig:hs_high}); about half a percent of the 10,000 scanned points are allowed by all direct search constraints but excluded by HiggsSignals.

\begin{figure}[!tbp]
\resizebox{0.5\textwidth}{!}{\includegraphics{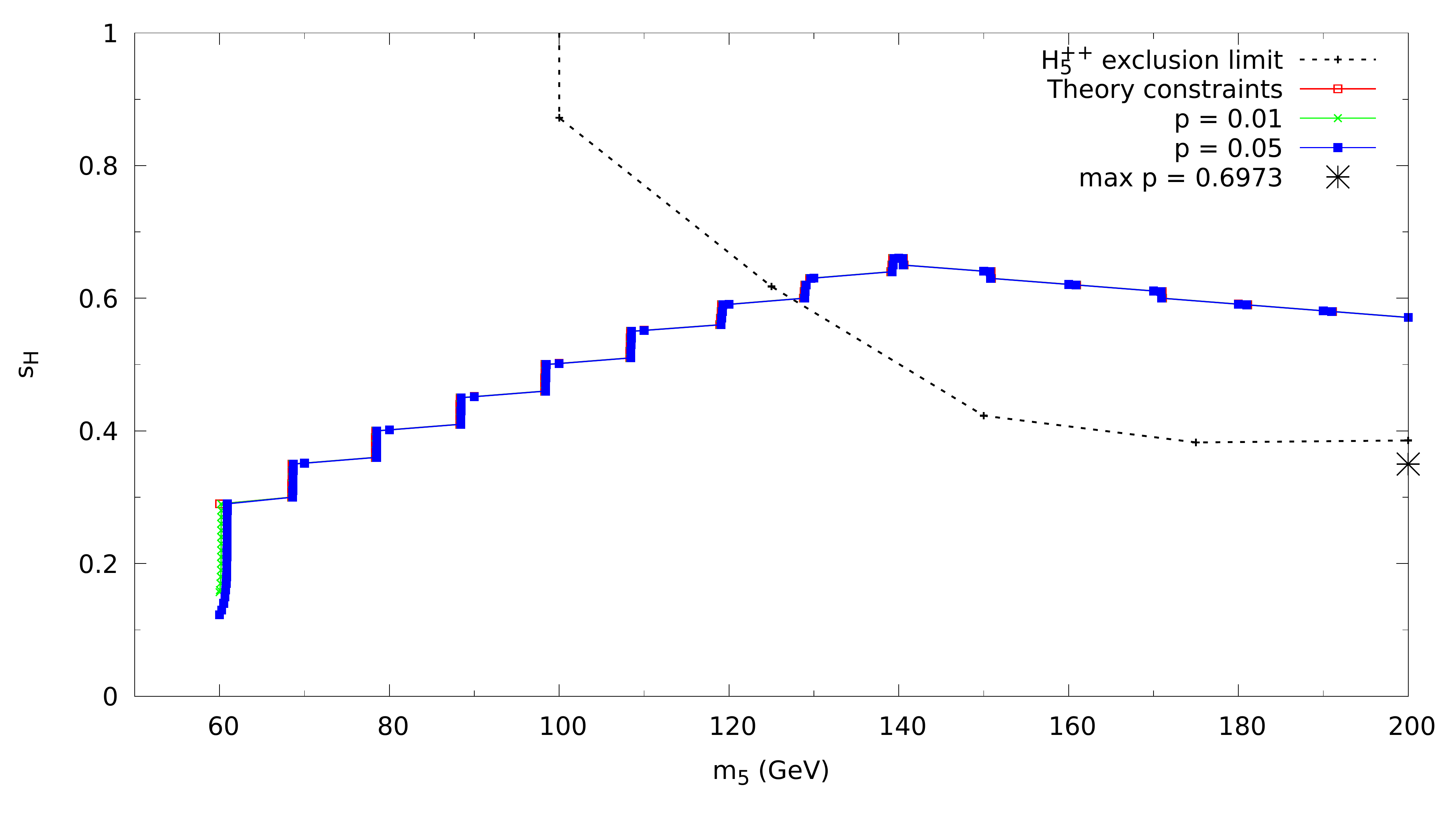}}%
\resizebox{0.5\textwidth}{!}{\includegraphics{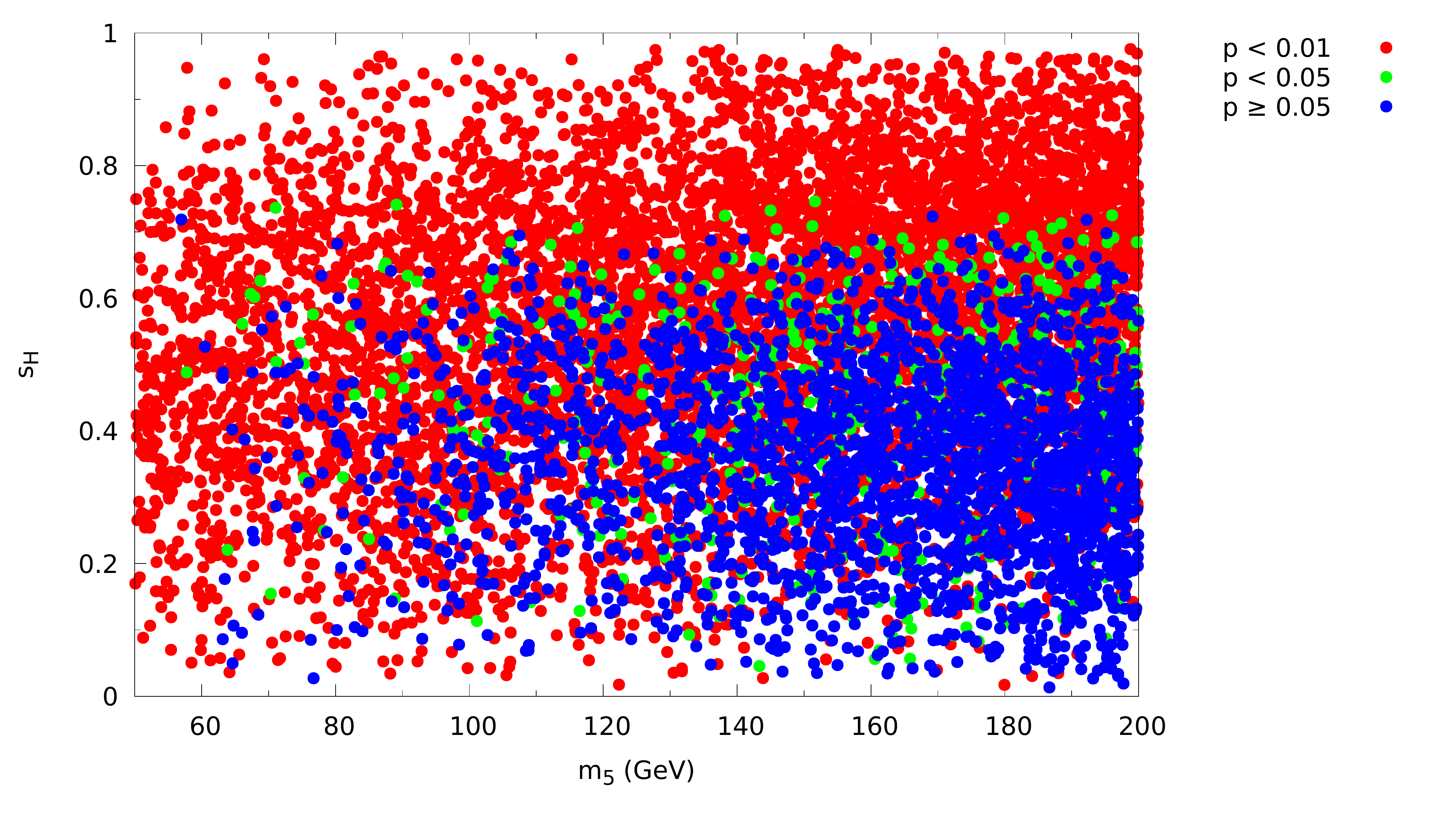}}
\caption{Left: contour (following the boundary of the theoretically-allowed region) of the $p$-value from a fit to experimental measurements of the 125~GeV Higgs boson's couplings in the low-$m_5$ benchmark calculated with HiggsSignals. The region above the black dotted line is excluded by an ATLAS constraint on VBF production of $H_5^{\pm\pm}$ decaying to $W^{\pm} W^{\pm}$, described in~\autoref{sec:constraints}. Right: as in the left panel but for a general scan of 10,000 points in the GM model with $m_5 \in [50, 200)$ GeV.}
\label{fig:hs_low}
\end{figure}

The constraints on the low-$m_5$ benchmark from HiggsSignals are shown in~\autoref{fig:hs_low} (left); in this benchmark the $p = 0.05$ contour follows the boundary of the theoretically-allowed region, meaning that all of the parameter space shown is allowed at the 95\% confidence level by the measured Higgs signal strengths.  This is in part due to the fact that the low-$m_5$ benchmark was designed so that the contribution of $H_5^+$ and $H_5^{++}$ to the loop-induced $h \to \gamma\gamma$ is suppressed.  The maximum $p$-value in this benchmark is 0.70.  The picture in the general scan is more complicated, with excluded points throughout the $s_H$ and $m_5$ range considered (right panel of \autoref{fig:hs_low}); about 11\% of the 10,000 scan points are allowed by all direct search constraints but excluded by HiggsSignals.  There are, however, also a large number of allowed points with $m_5 \gtrsim 60$~GeV and values of $s_H$ up to 0.7.  

In particular, the constraints on the parameter space of the GM model from the measured signal strengths of the 125~GeV Higgs boson are in general less constraining than those from direct searches for the new Higgs bosons.


\section{Discussion and conclusions}
\label{sec:conclusions}

In this work we evaluated constraints on the Georgi-Machacek model from direct searches for additional Higgs bosons and from signal strength measurements of the 125~GeV Higgs. To apply these constraints we used the general-purpose codes HiggsBounds 5.2.0 and HiggsSignals 2.2.1.  A few processes not included in HiggsBounds are particularly constraining for the GM model, which we implemented directly: these comprise searches for doubly-charged Higgs bosons and for Drell-Yan production of $H_5^0 H_5^{\pm}$ with $H_5^0 \to \gamma\gamma$.  We examined these constraints in the context of the H5plane and low-$m_5$ benchmark scenarios, as well as in two general scans over the full 7-dimensional parameter space of the GM model with $m_5$ ranges matched to those of the benchmarks.  Large regions of parameter space remain allowed by the current data.

One important result of our application of HiggsBounds to the GM model is to highlight the potential constraining power of the $H_3^0 \rightarrow Zh$ and $H \rightarrow hh$ channels at the LHC.  These already exclude small regions of parameter space that are not constrained by any other search in the GM model (as most clearly seen in the full scan for $m_5 > 200$~GeV).  Improvements in the sensitivity of these searches are thus important to further constrain the GM model and could lead to a discovery. 

\begin{figure}[!tbp]
    \includegraphics[width=0.8\textwidth]{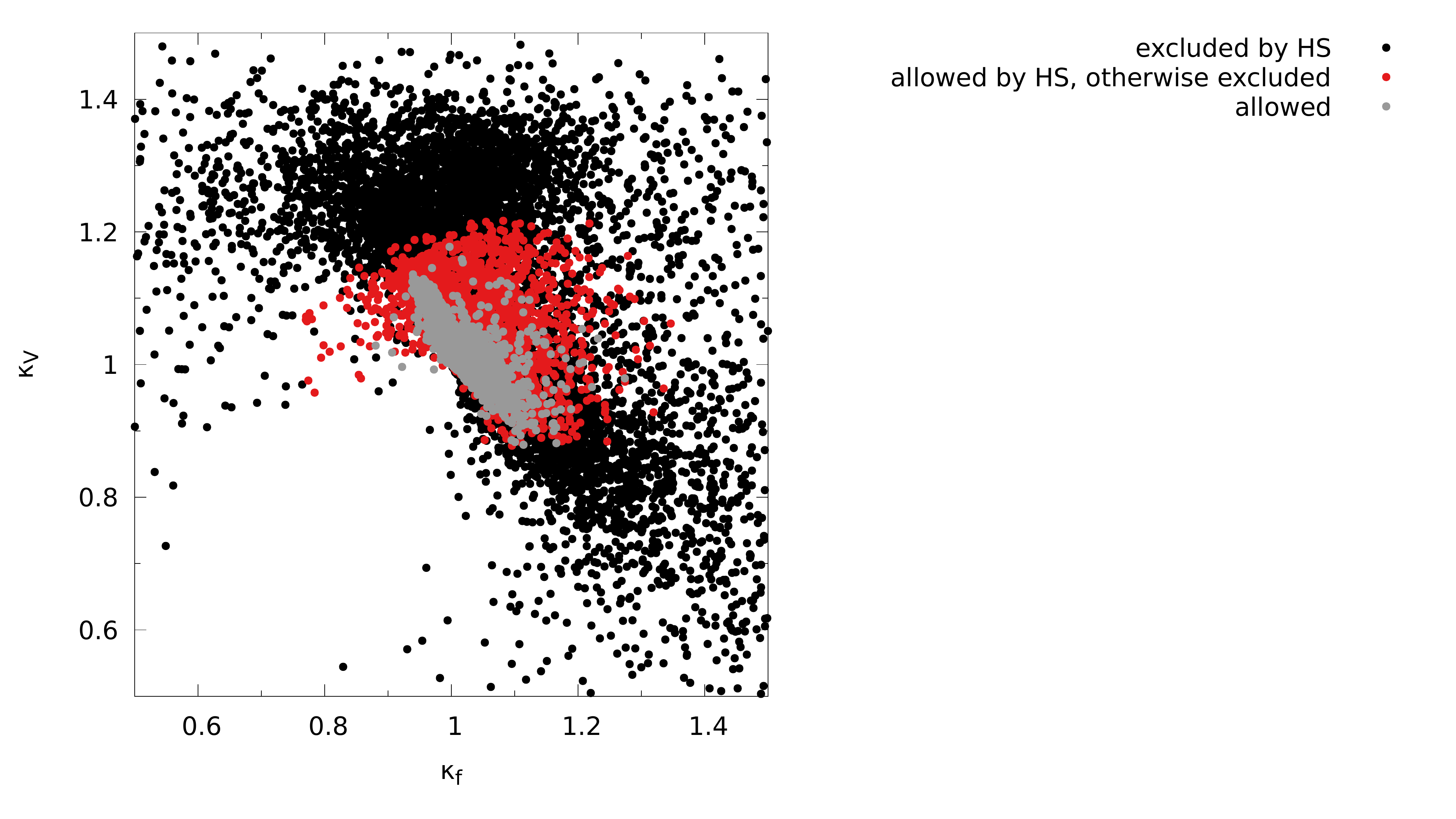}
    \caption{Allowed values of $\kappa_f$ and $\kappa_V$ (gray points) for the 125~GeV Higgs boson after applying constraints from HiggsSignals, HiggsBounds, and the direct searches that we implemented directly in GMCALC.  Points allowed by HiggsSignals but excluded by direct searches for additional Higgs bosons are shown in red.  Points from both the high- and low-mass general scans are included. $\kappa_f$ and $\kappa_V$ are defined as the $hf \bar f$ and $hVV$ couplings normalized to their values in the SM. Some excluded points lie beyond the boundaries of the plot.}
    \label{fig:summary}
\end{figure}

A second, perhaps counterintuitive, result is the relative weakness of the constraints from 125~GeV Higgs boson coupling measurements, as implemented via HiggsSignals.  In particular, direct searches for the additional Higgs bosons constrain the model parameter space in such a way as to limit the tree-level couplings of the 125~GeV Higgs bosons to fermions and to vector boson pairs to lie in the region $\kappa_f \in (0.88, 1.28)$ and $\kappa_V \in (0.87, 1.18)$ (\autoref{fig:summary}), where $\kappa_f$ and $\kappa_V$ are defined as the $hf \bar f$ and $hVV$ couplings normalized to their values in the SM.  Indeed, the fact that the direct searches for the additional Higgs bosons generally provide the more stringent constraints on the GM model parameter space than the 125~GeV Higgs boson signal strengths do leaves open the possibility of a discovery as these searches are improved with additional LHC data.

The code written to interface the model calculator GMCALC with HiggsBounds and HiggsSignals will be included in GMCALC versions 1.5.0 and higher, allowing the constraints to be easily updated as new experimental results are incorporated into HiggsBounds and HiggsSignals.

\begin{acknowledgments}
H.E.L.\ thanks Sakina Hussein for collaboration during initial investigations of the $H \to hh$ channel and Otto Eberhardt and Jana Schaarshmidt for fruitful discussions.  
This work was supported by the Natural Sciences and Engineering Research Council of Canada.  
A.I.\ was also partially supported by a Cornell Presidential Life Science Fellowship.
H.E.L.\ was also partially supported through the grant H2020-MSCA-RISE-2014 no.\ 645722 (NonMinimalHiggs). 
\end{acknowledgments}

%

\bibliography{references}
\end{document}